\documentclass[pre,twocolumn,showpacs,preprintnumbers,floatfix]{revtex4}
\usepackage{graphicx,citesort,psfrag,amsmath,amssymb}

\renewcommand{\vec}[1]{\mathbf{#1}}
\newcommand{\avg}[1]{\left\langle #1\right\rangle}
\newcommand{\mint}[4]{\int_{#2}^{#3}\!\!#1\,#4}

\newcounter{fnt}

\begin{document}
\bibliographystyle{apsrev}

\title{Overdamped Stress Relaxation in Buckled Rods} \author{Oskar
  Hallatschek} \author{Erwin Frey} \author{Klaus Kroy}
\affiliation{Hahn--Meitner Institut, Glienicker Stra\ss e 100, D-14109
  Berlin, Germany} \affiliation{ Fachbereich Physik, Freie Universit\"at
  Berlin, Arnimallee 14, D-14195 Berlin, Germany }

\date{\today}

\begin{abstract}
  We present a comprehensive theoretical analysis of the stress
  relaxation in a multiply but weakly buckled incompressible rod in a
  viscous solvent. In the bulk two interesting regimes of generic
  self--similar intermediate asymptotics are distinguished, which give
  rise to two classes of approximate and exact power--law solutions,
  respectively. For the case of open boundary conditions the
  corresponding non--trivial boundary--layer scenarios are derived by
  a multiple--scale perturbation (``adiabatic'') method.  Our results
  compare well with --- and provide the theoretical explanation for
  --- previous results from numerical simulations, and they suggest
  new directions for further fruitful numerical and experimental
  investigations.
\end{abstract}




\maketitle

\section{Introduction}

Any child that has played with a ruler during a boring school lesson,
has experienced the diverting physics of the paradigm of a mechanical
instability: The sudden buckling of a slender rod under a compressive
axial load of weight $f$ surpassing the first critical Euler force
$f_1$.  This so called Euler buckling instability is not only a
well--known example of a simple mechanical system exhibiting
non--trivial elastic behavior, historically it is also associated with
the birth of bifurcation theory. Its thorough understanding can temper
our intuition as to what should be expected or searched after in more
complicated situations involving elastic instabilities or bifurcations
in general. Intriguingly, it has also proved to be of major importance
for the equilibrium thermodynamic properties of stiff biopolymers
\cite{wilhelm-frey:96,legoff-hallatschek-frey:02}, such as actin or
collagen, which are largely responsible for the elastic properties of
biological tissue.  Recently also the dynamics of the Euler
instability has gained considerable interest as one of the most
elementary elastohydrodynamic problems \cite{wiggins-etal:98}. The
latter are commonly encountered in the derivation of macroscopic
constitutive models for soft, viscoelastic materials, i.e.\ materials
that show a mixed elastic and viscous behavior. For major examples of
this important type of condensed matter, ranging from polymer
solutions and gels to biological cells, the complicated dynamic
response can indeed be attributed to the elastohydrodynamics of some
low--dimensional meso--scale structures. Thus the focus has shifted
away from the classical treatment of the Euler instability
\cite{landau-lifshitz-7}, which is motivated by typical engineering
problems such as the stability of a mechanical beam under compressive
loads, to thermally undulated rods.  A crucial difference between the
two situations is that usually only the first few Euler modes matter
in the former, whereas (infinitely) many modes are excited in the
latter.

\begin{figure}
  \psfrag{ri}{$R(t=0)\lesssim L$} \psfrag{rf}{$R(t\to\infty)=L$}
  \psfrag{time}{time}
 \begin{center}
   \includegraphics[width=\columnwidth]{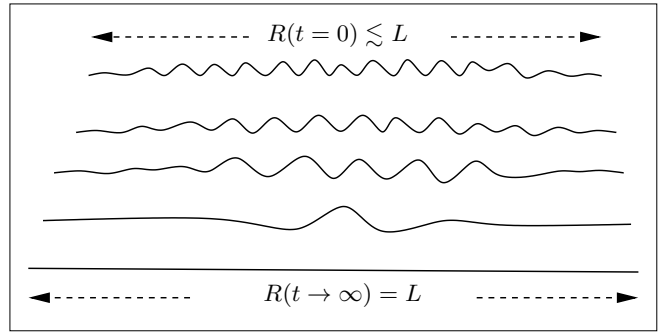}
\caption{A typical scenario of a deterministically relaxing buckled rod:
  Initially the rod is wrinkled on small wavelengths. In the course of
  time undulations are pushed out at the free ends and the typical
  wavelength of the undulations grows.}\label{fig:setting}
\end{center} 
\end{figure}

In the present contribution we are interested in deterministic
(``athermal'') dynamics under circumstances where many modes
contribute. Despite this restriction, our methods and major results
are also pertinent to certain ``thermal'' problems.  A telling example
is provided by the successful application of scaling arguments based
on deterministic dynamics to rationalize the non--equilibrium response
of a semiflexible polymer suddenly pulled at one end
\cite{seifert-wintz-nelson:96,ajdari-juelicher-Maggs:97}. More
precisely, we will consider here the deterministic overdamped
relaxation of the tension in an incompressible buckled rod as
schematically depicted in Fig.~\ref{fig:setting}. Initially, the
contour is strongly wrinkled on short length scales causing the
end--to--end distance $R(t=0)\lesssim L$ to slightly deviate from the
contour length $L$. It then evolves in time towards a completely
straight final state $R(t\to\infty)=L$ by transferring contour length
``stored'' in the high Euler modes to successively lower modes with
fewer and fewer nodes. The elastic energy stored in the compressed
initial state is thereby dissipated to the solvent.  The athermal case
already exhibits a very rich phenomenology (emergence of a
characteristic wavelength, exact and approximate power--law
relaxation, helix formation, staircase relaxation), only some of which
has previously been observed in numerical simulations
\cite{golubovic-moldovan-peredera:98,spakowitz:2001}.  These earlier
studies also provided scaling arguments rationalizing some of the
observations on the basis of a mathematical description adapted to the
simulation technique, which involves a compressible rod. In contrast,
our analysis starts from the mathematical minimal model \footnote{In
  contrast to related studies on $2D$ membranes
  \cite{moldovan-golubovic:99}, where a finite
  ``backbone''--compressibility is in fact a necessary ingredient, it
  does not play a vital role for the various phenomena of interest in
  the present contribution.} for the various phenomena of interest
outlined above, which is a contour $\vec r(s,t)$ parametrized by its
arc length $s=0\dots L$ and subject to an energetic cost
\begin{equation}\label{eq:wlc}
 {\cal H}[\vec r(s)] = \frac\kappa2\mint{ds}{0}{L}{\vec r''(s)^2}
\end{equation}
for bending that is proportional to the square of the local curvature
$\vec r''(s)$ (where we have introduced the shorthand notation $\vec
r'\equiv \partial\vec r/\partial s$).  The local incompressibility of
the contour has to be imposed onto Eq.~(\ref{eq:wlc}) as an external
rigid constraint
\begin{equation}\label{eq:rigid}
\vec r'(s)^2=1\;,
\end{equation}
which considerably complicates the calculations compared to classical
polymer models with fluctuating contour length \cite{doi-edwards:86}.
For finite temperatures, this model is generally known as the
Kratky--Porod model or wormlike--chain model in the polymer literature
\cite{doi-edwards:86,yamakawa}. However, as we said, here we focus on
its deterministic (zero--temperature) dynamics, exclusively.  The
contour is embedded into a highly viscous solvent of viscosity $\eta$,
and in the low Reynolds--number and free--draining limit one
approximates the viscous friction (per length) by two coefficients
$\zeta_\perp= 2\zeta_\| \approx 4\pi\eta$ for transverse and
longitudinal motion relative to the solvent, respectively
\cite{doi-edwards:86}.

We emphasize that a crucial ingredient implicit in related earlier
studies, is the weakly--bending--rod limit. It asserts that the local
slope of the contour is small. This condition has to be met for a
large negative line tension (pressure) $f\gg f_1$ to build up along
the contour. (Formally $f$ plays the role of a Lagrange multiplier
enforcing the incompressibility constraint onto $\cal H$.) Moreover,
the condition of weak bending naturally provides a small parameter
\begin{equation}\label{eq:wbrl}
 \epsilon=1-R/L\ll 1 \;,
\end{equation}
the fraction of the contour length ``stored'' in the contour
undulations.  Technically, the existence of this small parameter is
vital for the analytical approach to the problem. It enables us to
establish two independent mechanisms behind the ubiquitous
\cite{golubovic-moldovan-peredera:98,spakowitz:2001} power--law
temporal decay of the tension
\begin{equation}\label{eq:ft}
f(t)\propto t^{-1/2} \;.
\end{equation}
We will show that Eq.~(\ref{eq:ft}) may generically emerge either as a
consequence of a self--amplifying peak structure in the Fourier
decomposition of the contour $\vec r(s,t)$ or from self--affine
spatial correlations in the initial conditions $\vec r(s,0)$. It will
turn out that in the first case the structural relaxation proceeds
hand in hand with tension relaxation (\emph{type~I} behavior), whereas
in the second case it occurs essentially stress--free, after the
tension has already relaxed (\emph{type~II} behavior). In both cases,
we will also derive the associated growth laws for the boundary
layers near free ends and analyze their contribution to the relaxation
of the rod. The required adiabatic method of slowly varying tension,
which we develop in Section~\ref{sec:adiabatic_approximation} and in
the appendix~\ref{sec:homogenization}, can be generalized to
stochastic dynamics \cite{hallatschek-frey-kroy:04} and thus provides
a conceptional basis for a unifying description of tension propagation
in slender rods. The scenarios established for the tension relaxation
entail corresponding power--law scenarios for a number of observables
such as the dissipated energy or the growth of the radius of gyration
or end--to--end distance, which will be compared to simulations where
available.

The remainder is organized as follows.  In the next section, we
further specify the problem and give some intuitive arguments as to
its mathematical structure and the expected dynamics. For those
readers who happen to be mainly interested in a qualitative overview
over the rich deterministic dynamics of the Euler instability, we
moreover give a comprehensive qualitative and phenomenological
discussion of the results. Sec.~\ref{sec:qual} can also be read as an
extensive introduction to and outline of the detailed calculations and
results reported in the subsequent sections and in the appendix.

\section{Qualitative discussion}\label{sec:qual}

It is a remarkable feature of the classical analysis of the statics of
beam buckling that for the extremely non--linear problem of Euler
buckling a linear stability analysis is sufficient to determine the
onset of buckling. More precisely, after decomposing the rod contour
into discrete Fourier modes with amplitudes $a_n$, it yields the
associated critical forces $f_n=\kappa(\pi n/L)^2$ (here for the case
of hinged ends) necessary to excite these modes
\cite{landau-lifshitz-7}. The corresponding bending energies as a
function of the end--to--end distance $R$ follow as ${\cal
  H}(R)=f_n L\epsilon$ in the weakly bending limit.  This suggests that
also the \emph{dynamics} of the instability should be accessible to an
essentially linear calculation for a weakly bending rod, although the
problem outlined above is intrinsically strongly nonlinear. We will
show below that this is indeed the case as long as the tension along
the rod is sufficiently uniform. Then the dynamics can be understood
as arising from a linear superposition of relaxing eigenmodes that are
only \emph{globally} coupled by the incompressibility constraint
Eq.~(\ref{eq:rigid}). The latter restrains exponential growth of the
unstable modes by selecting the intermediate asymptotic power--law
relaxation Eq.~(\ref{eq:ft}) of the tension $f(t)$.

\begin{figure}
  \psfrag{g}{ground state} \psfrag{e}{excited state}
  \psfrag{1}{$r=(1-\epsilon)l$} \begin{center}
    \includegraphics[width=\columnwidth]{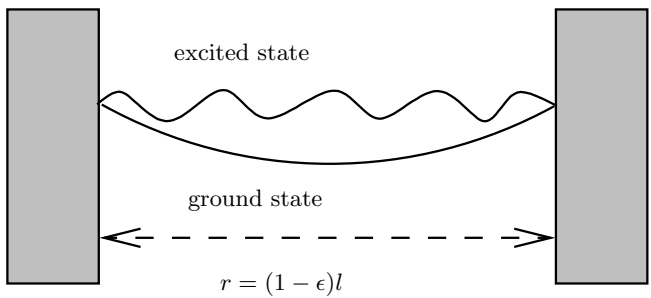}
 \caption{Relaxation in the bulk: The situation is essentially the
   same as for a longitudinally confined rod. The pressure $f(t)$
   exerted onto the confining walls exhibits the power--law decay
   Eq.~(\ref{eq:ft}).}\label{fig:bulk_pictogram} \end{center}
\end{figure}

Although our mathematical analysis applies more generally, it is
instructive to take the example of a free rod with a special initial
condition as a starting point. Namely a contour, which is wrinkled at
short scales with wrinkles that are statistically uniform along the
whole contour.  Obviously the relaxation at the free ends will not be
the same as in the bulk, but for the time being, we concentrate on the
bulk behavior. Take an arbitrary chosen short segment of length $l$
far away from the rod ends. To fully relax its bending energy it would
have to release its stored length $\epsilon\, l$ and thus to expand.
To this end, the sections of the rod to both of its sides would have
to be pushed out. Since these were assumed to be very long and almost
straight, so that their displacement is subject to substantial viscous
friction from the embedding fluid, this is virtually impossible for a
considerable period of time. (Note that the assumption of an almost
straight contour is crucial at this point.) The chosen initial
condition therefore entails that a uniform axial pressure $f$ much
larger than the critical pressure $f_1$ for the ground state builds up
along the contour. For a first analysis we may therefore imagine the
chosen \emph{bulk section} of the rod to be caged between two immobile
boundaries of distance $r=(1-\epsilon) l$ that preserves the total
stored length
\begin{equation}\label{eq:el}
\epsilon l=l-r = \text{ constant,}
\end{equation}
as depicted in Fig.~\ref{fig:bulk_pictogram} and analyzed in
Sec.~\ref{sec:lin}. The initial pressure $f(t=0)$ within the section
(which is the negative of the force exerted onto the boundaries) can
however still relax by transferring stored length from high modes to
low modes. If, for the sake of the argument, we imagine the initial
conformation to have essentially the form of a sine function with a
very small wavelength. In other words, mode amplitudes $a_n(0)$ shall
be peaked around some $N\gg1$ in mode space, say all $a_{n<N}$ are
extremely small and all $a_{n>N}$ vanish identically. Then the initial
pressure $f_N$ is much higher than the final (ground state) pressure
$f_1$. It will therefore relax by transferring the conserved stored
length from the $N^{\rm th}$ mode to successively lower modes, thereby
dissipating stored elastic energy to the solvent.  Since lower modes
have longer relaxation times they evolve more slowly, and the transfer
of stored length $\epsilon l$ happens via a cascade involving all
intermediate modes.  It turns out that the initial localization of
stored length and bending energy in mode space is not lost. The
numerical solution in Section~\ref{sec:cascade} will explicitly
confirm that under such conditions the transfer actually occurs in a
discontinuous jump mode leading to a staircase relaxation of the line
tension $f(t)$ and the corresponding confinement force around the
power law Eq.~(\ref{eq:ft}). Further, we will demonstrate analytically
that the localization in mode space emerges asymptotically from a
certain type of initial conditions, which are classified as
\emph{type~I} initial conditions. We will discuss in detail how the
global coupling of the modes via the constrained end--to--end distance
$r$ selects (up to logarithmic corrections) the power--law decay
Eq.~(\ref{eq:ft}) as intermediate asymptotics. It will be shown that
the localization in mode space consecutively sharpens with time,
thereby establishing the above mentioned staircase relaxation as a
generic long--time feature of \emph{type~I} behavior.  The mechanism
behind the localization in mode space will be seen to be formally
analogous to the onset of phase separation after a deep quench, i.e.\ 
to the early stages of spinodal
decomposition~\cite{cahn-hilliard:58,bray:94}.

A more thorough analysis of the initial conditions giving rise to
power--law relaxation in the bulk will be performed in
Section~\ref{sec:similarity_solutions}. The key observation is that
apart from the just mentioned intermediate asymptotics there is a
second class of similarity solutions \emph{exactly} obeying the
scaling behavior Eq.~(\ref{eq:ft}). They correspond to \emph{type~II}
initial conditions characterized by a power--law distribution of the
initial mode amplitudes. Among these is the particularly interesting
``thermally'' undulated contour (the dynamics still supposed to be
athermal).  Contrary to \emph{type~I} behavior, where
Eq.~(\ref{eq:ft}) can be understood as immediate consequence of the
appearance of a time--dependent characteristic wavelength
\begin{equation}\label{eq:qt}
 Q^{-1}(t) \propto t^{1/4}
\end{equation}
that visibly dominates the contour undulations, no palpable dominant
length scale (and hence no generic staircase relaxation) develops for
this second class of solutions. In fact, hardly any conformational
relaxation is noticeable during the decay of the tension for
\emph{type~II} initial conditions and the structural dynamics is
predominantly stress--free. Dynamic scaling can be attributed to the
self--affine geometry $a_n(0)\propto n^{-\beta/2-1}$ of the power--law
initial conditions, which are completely characterized by a
``\emph{roughness exponent}'' $\beta>1$.  The discussion in
Secs.~\ref{sec:cascade},~\ref{sec:similarity_solutions} will
eventually allow us to conclude that all \emph{generic}~\footnote{We
restrict our analysis to initial mode amplitudes that are either
peaked or distributed according to a power law (possibly with some
cut--off) in mode space. We discard as ``non--generic'' those initial
conditions where one either starts essentially in the ground state or
has multiple peaks, oscillations etc. in the mode spectrum. This
attitude is commonly adopted in related studies
\cite{seifert-wintz-nelson:96,ajdari-juelicher-Maggs:97}.} initial
conditions invariably give rise to the same universal power--law
relaxation Eq.~(\ref{eq:ft}) of the force but with variable degree of
localization in mode space, as summarized by
Fig.~\ref{fig:critical-alphas}.

\setcounter{fnt}{\value{footnote}}

While the discussion so far holds anywhere in the bulk of the rod,
where the longitudinal expansion can practically be neglected on the
appropriate logarithmic time scale, we will in the remainder also
address the slightly different situation near the free ends
(Section~\ref{sec:free-relaxation}). Surprisingly, it can be analyzed
along the same lines as the bulk by virtue of a length scale
separation innate to the weakly bending limit. The major variation of
the tension, namely from its bulk value to zero at the open
boundaries, occurs within a (time--dependent) \emph{boundary layer} of
length $\lambda(t)$ that is at any time much larger than the
characteristic length scale $Q^{-1}(t)$ of the dynamically most active
contour undulations.  This fortunate situation is schematically
depicted in Fig.~\ref{fig:coarse-graining}. It allows the derivation
of closed equations for the (smooth) coarse--grained tension profile
by means of an adiabatic approximation that integrates out the
contingent short wavelength fluctuations up to a coarse--graining
length scale $l(t)$ intermediate between $Q^{-1}(t)$ and $\lambda(t)$.
The underlying idea goes back to
Ref.~\cite{seifert-wintz-nelson:96}. As an aside, we point out a
subtle technical difference between the bulk and the boundary layer
problem, here. While the former is accessible to an ordinary (regular)
perturbative approach, the adiabatic approach to the latter amounts to
a multiple--scale perturbation scheme. The additional effort is
rewarded by the possibility to generalize the approach to arbitrary
situations that exhibit a slow (compared to $l$) ``systematic''
variation of the line tension and stored length along the contour.
The corresponding formalism is developed in
Section~\ref{sec:adiabatic_approximation} and in the
appendix~\ref{sec:homogenization} and enables us to derive the central
Eq.~(\ref{eq:coarse-graining}). It can be interpreted as a continuity
equation for the (coarse--grained) local stored length, which
generalizes Eq.~(\ref{eq:el}) to situations with spatially varying
tension $f(s)$. The application to the situation near the free ends
allows us to infer a non--trivial dynamic scaling laws for the
boundary layer. Its width $\lambda$ is found to grow according to
\begin{equation}\label{eq:lamba}
 \lambda(t)\propto t^\delta \;.
\end{equation}
The exponent $\delta$ characterizing this growth depends on the degree
of localization of the stored length in mode space, so that one again
has to distinguish between \emph{type~I} and \emph{type~II} behavior.
For \emph{type~I} initial conditions we find $\delta=1/4$, hence the
boundary layer width is proportional to (though numerically much
larger than) the wavelength of the dominant mode, i.e.\
$\lambda\propto Q^{-1}$. It thus does not represent a new
characteristic dynamic length scale itself. As we noticed for the
bulk, tension propagation and contour relaxation proceed in parallel.
Asymptotically the rod contour can be decomposed into a bulk region
with homogeneous line tension $f(s)=$ const.\ and two virtually
stretched end sections where the tension has relaxed to the linear
profile $f(s)\propto \eta s$ characteristic of a rigid rod subject to
a viscous friction force. These predictions compare well with the
available numerical simulations \cite{spakowitz:2001}. In contrast,
for \emph{type~II} initial conditions, which were not yet studied in
simulations, $\delta=(3-\beta)/8$ is predicted to depend on the
roughness exponent $\beta$, so that $\lambda$ provides a new
($\beta-$dependent) characteristic dynamic length scale besides
$Q^{-1}$. The vanishing of $\delta$ for $\beta\to3$ heralds the
(trivial) limit of instant equilibration. For $1<\beta<3$ tension
propagation precedes contour relaxation, so that most of the contour
relaxation occurs under vanishing tension. Such curious dependence of
\emph{type~II} relaxation behavior on the value of the exponent
$\beta$ was previously noticed in a different context
\cite{ajdari-juelicher-Maggs:97}. As an important special case, we
obtain the exponent $\delta=1/8$ for ``thermal'' initial conditions,
which coincides with the corresponding exponent for the
non--equilibrium thermodynamic tension propagation known from linear
response calculations \cite{everaers-Maggs:99}.

The divergence of tension decay (or propagation) and conformational
relaxation for \emph{type~II} initial conditions raises the question
how under these circumstances tension propagation can be observed in
experiments or simulations. In Sec.~\ref{sec:simulation} we establish
that by virtue of the general relation
\begin{equation}\label{eq:dtrg}
 \partial_t R_{G\parallel}(t) \propto f(t) \propto t^{-1/2} \;,
\end{equation}
the growth of the longitudinal component $R_{G\parallel}$ of the
radius of gyration is a suitable observable to directly monitor the
decay law Eq.~(\ref{eq:ft}) for the tension $f(t)$. This said,
$R_{G\parallel}$ obviously should not be regarded as a genuine measure
of the conformational dynamics. The latter can instead be accessed via
measuring the change $\delta R_\parallel(\tau)$ of the longitudinal
component of the end--to--end distance. Similarly as the boundary
layer width $\lambda$, it portraits the richer conformational dynamics
in its power--law growth
\begin{equation}\label{eq:dr}
 \delta R_\parallel(t)\propto t^\rho \;.
\end{equation}
Under \emph{type~I} conditions the exponent turns out to be $\rho=1/4$
as $\delta R_\parallel\propto \lambda\propto Q^{-1}$ is just a small
constant fraction of the boundary layer width.  On the other hand, we
predict a crossover from $\rho=(1+\beta)/8=\delta+(\beta-1)/4$ for
short times to $\rho=(\beta-1)/4$ for long times under \emph{type~II}
conditions, so that in this case, $Q^{-1}$, $\lambda$, and $\delta
R_\parallel$ all constitute different (albeit related) dynamic length
scales.

The above qualitative discussion has hopefully convinced the reader
that the dynamics of the mechanical Euler buckling instability
exhibits a rich and interesting phenomenology that deserves a more
detailed mathematical analysis. This is what the following sections
intend to provide.

\section{Equations of Motion}
\label{sec:EOM}

\begin{figure}[ht]
  \psfrag{x}{$s-r_\parallel$} \psfrag{y}{$r_{\perp ,1 }$}
  \psfrag{z}{$r_{\perp ,2}$} \psfrag{L}{$L$} \psfrag{0}{$0$}
\begin{center}    
  \includegraphics[scale=.9]{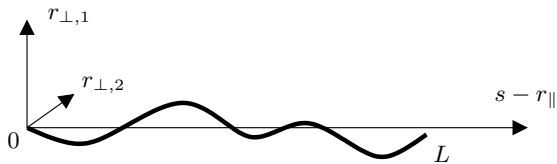}
    \caption{The parameterization of the contour $\vec r(s)=(\vec
      r_\perp,s-r_\parallel)^T$ by transverse and longitudinal
      displacement variables, $r_\perp$ and $r_\parallel$,
      respectively. Note, that the displacements vanish for the
      straight contour.}
    \label{fig:parametrization}
  \end{center}
\end{figure}

As motivated in the Introduction, the axial incompressibility can lead
to a large negative tension (pressure) in a relaxing rod. This crucial
feature only appears for almost straight rods or straight rod
sections. We therefore concentrate on the geometry of an almost
straight rod and introduce displacement variables that describe the
deviation from the straight contour, as depicted in
Fig.~\ref{fig:parametrization}.
We parametrize the contour by $\vec r=(\vec r_\perp,s-r_\parallel)^T$,
where $\vec r_\perp(s)$ is the two--dimensional transverse
displacement vector at arclength $s$ and
$r_\parallel(s)-r_\parallel(0)$ is the contour length stored in
undulations within the rod section $(0,s)$. For being almost straight
the contour has to have small transverse slope at any given point,
i.e.
\begin{equation}
  \label{eq:wkb}
  r_\perp'^2(s,t)\ll1 \qquad \mbox{(weakly bending limit)}\;.
\end{equation}
The inextensibility of the rod Eq.~(\ref{eq:rigid}) couples transverse
and longitudinal coordinates. Resolving it for $r_\|'$ and expanding
the square root, it reads
\begin{equation}
  \label{eq:local-const}
  r_\parallel'=\frac12\vec r_\perp'^2+O(r_\perp'^4) \;.
\end{equation}
Since the dimensionless quantity $r_\parallel'(s,t)\ll1$ will act as
the small parameter throughout the following treatment, we reserve the
variable $\epsilon(s,t)$ for it,
\begin{equation}
  \label{eq:epsilon}
  \epsilon(s,t)\equiv r_\parallel'(s,t) \ll1 \;.
\end{equation}
The function $\epsilon(s,t)$ can be interpreted as the fraction of the
contour length stored in the transverse undulations. Note that from
Eq.~(\ref{eq:local-const}) $r_\perp'^2$ is of order $O(\epsilon)$ and
the terms neglected in Eq.~(\ref{eq:local-const}) are of order
$O(\epsilon^2)$.

We now turn to the derivation of the equations of motion in terms of
$\vec r_\perp$ and $r_\parallel$. In the case of low Reynolds numbers
the dynamics is determined by the balance of elastic, driving and
friction forces. The elastic force derives from
\begin{equation}
  \label{eq:eff-free-energy}
  {\cal R}={\cal H}-\frac{1}{2}  \mint{ds}{0}{L}  f\vec r'^2
\end{equation}
via functional differentiation \cite{goldstein-langer:95}. The
Lagrange multiplier function $f(s,t)$ is necessary to preserve the
arclength constraint Eq.~(\ref{eq:rigid}). It can be interpreted as a
(negative) local line tension.

For elongated slender bodies like thin rods or stiff polymers, it is
well justified to assume a local anisotropic friction force (free
draining limit). The anisotropy is due to the fact that the friction
coefficient $\zeta_\perp=2\zeta_\parallel$ (per unit length) of a
stiff rod moving perpendicular to its long axis is twice as large as
that for longitudinal motion. The force balance is given by the
expression $\partial_t \vec r=-\vec H\,\delta {\cal R}/\delta \vec r$
with a hydrodynamic mobility tensor $\vec H$ associated with the
anisotropic friction~\cite{wiggins:03}. To order $\epsilon$, it takes
the form
\begin{subequations}\label{eq:EOM}
\begin{align}
  \zeta_\perp \partial_t \vec r_\perp&=
  -\kappa\vec r_\perp''''-(f \vec r_\perp')' \label{eq:perp-eq-motion} \\
  \zeta_\parallel\partial_t
  r_\parallel-(\zeta_\perp-\zeta_\parallel)\vec r_\perp'
  \partial_t\vec r_\perp&=-\kappa r_\parallel''''+f'-(f
  r_\parallel')'\label{eq:parallel-eq-motion} \;.
\end{align}
\end{subequations}
The local anisotropy of the friction generates additional terms of
order $O(\epsilon^{3/2})$ that are neglected here. 
For a freely relaxing rod with given initial conditions, the equations
of motion Eqs.~(\ref{eq:EOM}) have to be solved respecting the local
constraint Eq.~(\ref{eq:local-const}) and the boundary conditions of
zero tension, torque and force at the ends,
\begin{equation}
  \label{eq:free-bc}
  f|_{0,L}=\vec r''|_{0,L}=\vec r'''|_{0,L}=0 \;.
\end{equation}

\section{Relaxation of a Confined Weakly Bending Rod}\label{sec:lin}
\subsection{The Leading Order in $\epsilon$}\label{sec:leading_order}

In course of our qualitative discussion in Sec.~\ref{sec:qual} we
motivated that the key--problem for understanding the bulk of a
relaxing rod is the relaxation of a rod section of length $l\ll L$
confined between two immobile walls, as illustrated in
Fig.~\ref{fig:bulk_pictogram}. The weakly bending rod section is
supposed to be initially perturbed by small wrinkles of wavelength
much smaller than $l$ (excited state).  Owing to the undulations, the
end--to--end distance $r$ is by an amount $\epsilon(t) l$ smaller than
the total length $l$, where
\begin{equation}
  \label{eq:wkb-condition}
  \epsilon(t) \equiv \frac{1}{l}\mint{ds}{0}{l}\epsilon(s,t)=
  \frac{r_\parallel(l)-r_\parallel(0)}{l}
\end{equation}
is the spatial average of the stored length density $\epsilon(s,t)$.
With the help of Eq.~(\ref{eq:local-const}), we can express
$\epsilon(t)$ in terms of the transverse displacements,
\begin{equation}
  \label{eq:global-constraint2} \epsilon(t)l
  =\frac{1}{2}\mint{ds}{0}{l}\vec r_\perp'^2(s,t)+O(\epsilon^2) \;.
\end{equation}

By exerting a compressing force $f(t)$ on the rod ends, the walls keep
the rod section from expanding and the total stored length
$\epsilon(t)l$ remains constant,
\begin{equation}
  \label{eq:global-constraint}
  \Delta \epsilon(t)\equiv\epsilon(t)-\epsilon^0 =0 \;,
\end{equation}
where $\epsilon^0=\epsilon(0)$ is the initially stored contour length.
Our question is: How does such an ``excited'' rod relax to the ground
state, in which the contour only has one buckle of wavelength $l$ (as
depicted in Fig.~\ref{fig:bulk_pictogram})?

In the present section we will apply regular perturbation theory to
address this problem, i.e. all derivations are understood to hold to
leading order in $\epsilon$. This allows us to neglect the spatial
dependence of the tension. The longitudinal equation of motion
Eq.~(\ref{eq:parallel-eq-motion}) together with Eq.~(\ref{eq:epsilon})
implies that
\begin{equation}
  \label{eq:f-prime-small} 
f'= O(\epsilon) \;.
\end{equation}
Therefore spatial variations of the tension are small in the limit
$\epsilon\ll1$ and the transverse equation of motion
Eq.~(\ref{eq:perp-eq-motion}) is to leading order $O(\sqrt{\epsilon})$
given by
\begin{eqnarray}
  \zeta_\perp\partial_t \vec r_\perp&=&-\kappa \vec r_\perp''''-f(t)
  \vec r_\perp'' \;, \label{eq:transverse-eom-lowestorder}  
\end{eqnarray}
where merely the spatial average
\begin{equation}
  \label{eq:spatial-force-average}
   f(t)\equiv \frac{1}{l}\mint{ds}{0}{l}f(s,t) \;
\end{equation}
of the force $f(s,t)$ enters. The longitudinal force that the walls
exert on the segment equals $f$ up to terms of order $\epsilon$.
Though Eq.~(\ref{eq:transverse-eom-lowestorder}) is linear for a given
force history, the global constraint of fixed end--to--end distance
Eq.~(\ref{eq:global-constraint}) makes the tension a functional of the
contour $\vec r_\perp(s,t)$. The resulting problem comprised by
Eqs.~(\ref{eq:global-constraint},~\ref{eq:transverse-eom-lowestorder})
is therefore still highly non--linear and in general not analytically
tractable. Progress can be made, however, for generic
cases~\footnotemark[\value{fnt}], as will be shown in the following
subsections. We will also present exact numerical solutions in order to
illustrate the results.

\subsection{Amplification Factor}
\label{sec:ampelfaktor}

We analyze the problem in two steps: For a given force history $f(t)$
Eq.~(\ref{eq:transverse-eom-lowestorder}) is linear in $\vec r_\perp$.
Therefore, we can determine the stored length as a function of the
tension via Eq.~(\ref{eq:global-constraint2}). The second (in general
non trivial) task is then to revert this relation and to determine the
correct force history, that obeys Eq.~(\ref{eq:global-constraint}) by
keeping the end--to--end distance $r$ constant.

We decompose the contour into sine functions,
\begin{equation}
  \label{eq:hinged-ends}
  \vec r_\perp(s,t)=\sqrt{2/l}\sum_n \vec a_n(t) \sin(q_n s) \;,
\end{equation}
where $q_n=n\pi/l$ is the wave number corresponding to the $n^{\rm
  th}$ mode, and for definiteness hinged ends have been assumed for
the boundary conditions. (The same boundary conditions have also been
used in the molecular dynamics simulations of
Ref.~\cite{golubovic-moldovan-peredera:98}.)  Then,
from Eq.~(\ref{eq:global-constraint2}) the stored length can be
written as
\begin{equation}
  \label{eq:stole-modespace}
  \epsilon(t) \, l=\frac{1}{2}\sum_n q_n^2 \vec a_n^2 \equiv \sum_n
  \epsilon_n \, l \;.
\end{equation}
The elements $\epsilon_n(t) \, l$ of the last sum can be interpreted
as the contour length stored in mode $n$ at time $t$. We obtain a
dynamical equation for $\epsilon_n(t)$ by first inserting the Fourier
decomposition Eq.~(\ref{eq:hinged-ends}) into
Eq.~(\ref{eq:transverse-eom-lowestorder}) and then multiplying the
resulting equation for the mode amplitudes with $\vec a_n q_n^2/(2l)$:
\begin{equation}
  \label{eq:eom-stored-length}
  \partial_{\tau}\vec \epsilon_n=2 \left[- q_n^4+\varphi(\tau)
    q_n^2\right]\epsilon_n \;. 
\end{equation}
Here we introduced a rescaled tension $\varphi\equiv f/\kappa$ and
time $\tau\equiv\kappa t/\zeta_\perp$, which have units of
length$^{-2}$ and length$^4$, respectively. Now, all variables of our
problem represent powers of lengths.  The dispersion relation
Eq.~(\ref{eq:eom-stored-length}) exhibits a stable and an unstable
band of modes, separated by the wave number $\sqrt{\varphi(\tau)}$.
Modes with larger wave numbers shrink exponentially, whereas the
others grow exponentially as a consequence of the competition of the
{\em restoring} bending force (dominating at large wave numbers) and
the {\em driving} compressing force (dominating at small wave
numbers).  This is formally analogous to spinodal decomposition. In
this context Eq.~(\ref{eq:eom-stored-length}) with $\kappa$ measuring
the surface tension and $f$ the curvature of the local free energy at
the central maximum, is known as the Cahn--Hillard
equation~\cite{cahn-hilliard:58}.

\begin{figure}
  \psfrag{x}{$x$} \psfrag{A}{$A(q)$} \psfrag{AA}{$A(q)$}
  \psfrag{q}{$q/Q$} \psfrag{a=2}{$\alpha=2$}
  \psfrag{a=.15}{$\alpha=0.15$} \psfrag{(a)}{$(a)$}
  \psfrag{(b)}{$(b)$} \psfrag{1}{$1$} \psfrag{22}{$\sqrt{2}$}
  \psfrag{50}{$50$}    
  \includegraphics[width=\columnwidth]{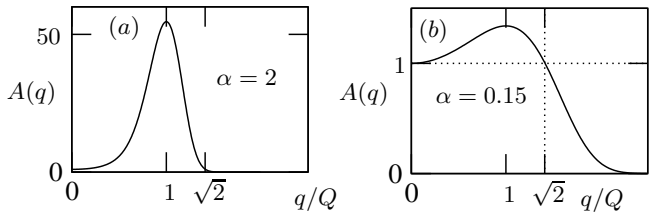}
    \caption{The sensitive dependence of $A\left(q,\tau\right)$,
      given by Eq.~(\ref{eq:ampl-scalingform2}), on the parameter
      $\alpha=\tau Q^4$. For $\alpha=2$ the amplification factor takes
      the form of a pronounced peak $(a)$, whereas for $\alpha=0.15$
      it resembles a step function.}
    \label{fig:greens-function}
\end{figure}

Integration of Eq.~(\ref{eq:eom-stored-length}) by separation of
variables yields
\begin{equation}
  \label{eq:greens_fct}
  \epsilon_n(\tau)=\epsilon^0_n A(q_n,\tau)
\end{equation}
with the initial values $\epsilon^0_n$ and an ``amplification factor''
\begin{equation}
  \label{eq:amplific-fac}
  A(q,\tau)\equiv \exp\left[2q^2 \left(\Phi(\tau)- q^2\tau\right)\right] \;,
\end{equation}
$\Phi$ being the time integral over the tension,
\begin{equation}
  \label{eq:Phi-def}
  \Phi(\tau)\equiv\mint{d\tilde \tau}{0}{\tau} \varphi(\tilde\tau) \;.
\end{equation}
The structure of the function $A(q,t)$ becomes more transparent upon
introducing the characteristic wave number $Q(\tau)$ corresponding to
the position of its maximum,
\begin{equation}
  \label{eq:characteristic-wnumber}
  Q^2(\tau)\equiv\frac{\Phi(\tau)}{2\tau} \;,
\end{equation}
related to $\varphi(\tau)$ by
\begin{equation}
  \label{eq:phifromQ}
  \varphi=\partial_\tau \Phi=2\partial_\tau(\tau Q^2) \;.
\end{equation}
Note that the wave number $Q(\tau)$ that has grown most strongly up to
time $\tau$ depends on the force history $\varphi(\tilde\tau<\tau)$.
With this definition, Eq.~(\ref{eq:amplific-fac}) is rewritten as
\begin{subequations}\label{eq:ampl-scalingforms}
\begin{align}
  A(q,\tau)&=\exp\left[2\tau q^2\left(2Q^2-q^2\right)\right]
  \label{eq:ampl-scalingform1} \\ &=\exp\left[2
  \alpha
  \left(q/Q\right)^2\left(2-\left(q/Q\right)^2\right)\right]\;,
  \label{eq:ampl-scalingform2}
\end{align}
\end{subequations}
where we introduced the dimensionless parameter 
\begin{equation}
  \label{eq:alpha-def} \alpha(\tau)\equiv\tau Q^4(\tau) \;.
\end{equation}
The amplification factor $A(q,\tau)$ describes how the stored length
is rearranged for a given force history. In general wave numbers
larger than $\sqrt{2}Q$ are damped ($A<1$) and wave numbers smaller
than $\sqrt{2}Q$ are amplified ($A>1$). Further the amplification
factor depends very sensitively on the parameter $\alpha$ defined in
Eq.~(\ref{eq:alpha-def}). For $\alpha\gg1$ the function $A(q,\tau)$
develops a strong peak around $q=Q$ with a height $e^{2\alpha}$ and a
relative width $\Delta Q/Q$ of about~\footnote{We chose $\Delta Q$
  somewhat arbitrarily to be twice the standard deviation of the peak.
  $\Delta Q^{-1}$ can be interpreted as a coherence length over which
  the contour of the rod can be considered to be a pure sinusoidal.}
\begin{equation}
  \label{eq:width}
  \Delta Q/Q=(4\sqrt{\alpha})^{-1/2}\;, 
\end{equation}
as shown in Fig.~\ref{fig:greens-function}(a), i.e.\ it somewhat
resembles the delta function $e^{2\alpha}\Delta Q\delta(q-Q)$. In the
case of $\alpha< 1$, which is illustrated in
Fig.~\ref{fig:greens-function}(b), the function resembles more the
step function $\Theta(\sqrt{2}Q-q)$.

We now turn to the second step of determining the force history
$\varphi(\tau)$, that makes the dynamics compatible with the
constraint of fixed end--to--end distance,
Eq.~(\ref{eq:global-constraint}). This is achieved by rewriting the
constraint Eq.~(\ref{eq:global-constraint}) inserting
Eqs.~(\ref{eq:stole-modespace},~\ref{eq:greens_fct},~\ref{eq:ampl-scalingform1}),
\begin{subequations}\label{eq:longitudinal-arrest}
  \begin{align}
  0 &=\Delta\epsilon(\tau) \\ &=\sum_n
  \epsilon^0_n\left[A(q_n,\tau)-1\right] \label{eq:constr-2} \\
  &=\sum_n \epsilon^0_n \left\{\exp\left[2\tau
  q_n^2\left(2Q^2(\tau)-q_n^2\right)\right]-1\right\} \;.
  \label{eq:force-cond-mode-space}
\end{align}
\end{subequations}
For a given time $\tau$ Eq.~(\ref{eq:force-cond-mode-space}) is an
implicit equation for the characteristic wave number $Q(\tau)$ which
by its definition Eq.~(\ref{eq:characteristic-wnumber}) is related to
$\Phi(\tau)$, the time integral over the tension.

\begin{figure}[ht]
  \psfrag{q}{$q$} \psfrag{eA}{$\epsilon_n(\tau)$} \psfrag{C}{$q_N$}
  \psfrag{Q2}{$\sqrt{2}Q(\tau_3)$} \psfrag{Q3}{$Q(\tau_3)$}
  \psfrag{(a)}{$\Delta\epsilon_-$} \psfrag{(b)}{$\Delta\epsilon_+$}
  \psfrag{t1}{$\tau_1$} \psfrag{t2}{$\tau_2$} \psfrag{t3}{$\tau_3$}
  \psfrag{dQ}{$\Delta Q(\tau_3)$}
\begin{center}    
  \includegraphics[width=\columnwidth]{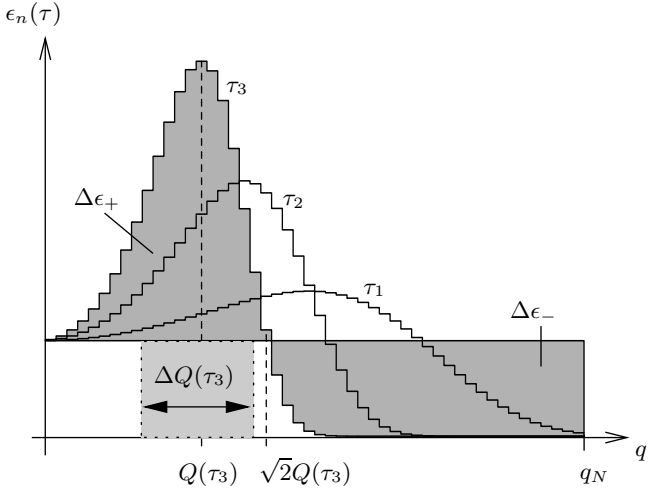} \caption{The
    fraction $\epsilon_n(\tau)=\epsilon^0_n A(q_n,\tau)$ of contour
    length stored in mode $n$ versus the corresponding wave number
    $q_n=n\pi/l$ for three successive times $\tau_3>\tau_2>\tau_1$ and
    the particular initial condition $\epsilon^0_{n\leq N}=\mbox{const.}$
    and $\epsilon^0_{n>N}=0$. The location of the maximum $Q(\tau)$ of
    $A(q,\tau)$ has been obtained upon solving the implicit equation
    (\ref{eq:force-cond-mode-space}) numerically. As explained in the
    main text the dark shaded areas $\Delta\epsilon_-$ and
    $\Delta\epsilon_+$ can be interpreted as stored length that has
    been destroyed and generated during the relaxation, respectively.
    The global constraint of fixed stored length requires their sum to
    vanish identically, Eq.~(\ref{eq:s+=s-}).}
    \label{fig:bow-wave} \end{center}
\end{figure}

The remainder of Sec.~\ref{sec:lin} is devoted to the analysis of the
time dependence of the solutions $Q(\tau)$ of
Eq.~(\ref{eq:force-cond-mode-space}). First of all, we note that
numerically it is straightforward to solve the implicit equation for
any initial condition $\epsilon^0_n$. This allows us to illustrate a
key feature of the relaxation process right away, namely the
continuous transfer of stored length from small to large
scales. Fig.~\ref{fig:bow-wave} shows the mode number dependent
fraction of stored length $\epsilon_n(\tau)$ at three successive times
$\tau_1<\tau_2<\tau_3$ for the initial condition
$\epsilon^0_{n\leq N}=\epsilon^0/N=\mbox{const.}$ and
$\epsilon^0_{n>N}=0$. For this particular choice of initial conditions
$\epsilon_n(\tau)$ can up to a constant prefactor be identified with
$A(q_n,\tau)$. The dark grey area represents the difference between
the stored length at time $0$ and time $\tau_3$. It has two natural
sub--divisions $\Delta\epsilon_+>0$ and $\Delta\epsilon_-<0$ adding up
to zero by virtue of the constraint Eq.~(\ref{eq:constr-2}). Formally,
we define by
\begin{equation}
  \label{eq:s-}
   \Delta\epsilon_-(\tau)\equiv
   \sum_{n=N_c}^{\infty}\epsilon^0_n\left[A(q_n,\tau)-1\right] <0 \;,
\end{equation}
with $N_c$ being the smallest $n$ with $n>\sqrt{2}Q(\tau)$, the stored
length that has been ``destroyed'' up to time $\tau$. Note, that each
element of the sum Eq.~(\ref{eq:s-}) is negative. Similarly,
$\Delta\epsilon_+(\tau)$ represents the stored length that has been
``generated'' in the modes with wave numbers below $\sqrt{2} Q(\tau)$,
i.e. we define
\begin{equation}
  \label{eq:s+}
   \Delta\epsilon_+(\tau)\equiv
   \sum_{n=0}^{N_c-1}\epsilon^0_n\left[A(q_n,\tau)-1\right] >0\;. 
\end{equation}
Since the total change in stored length $\Delta \epsilon$ must vanish,
\begin{equation}
  \label{eq:s+=s-} \Delta \epsilon(\tau)\equiv\Delta\epsilon_+(\tau) +
  \Delta\epsilon_-(\tau)=0 \;,
\end{equation} 
we can imagine the relaxation process as a \emph{transfer} of stored
length $|\Delta\epsilon_-(\tau)|$ from scales smaller than
$(\sqrt{2}Q)^{-1}$ to scales larger than $(\sqrt{2}Q)^{-1}$ in time
$\tau$.  Due to the form of the amplification function (see
Fig.~\ref{fig:greens-function}) the mode amplitudes with wave number
close to $Q(\tau)$ show the largest increase up to time $\tau$. In the
present case this results in the formation of a pronounced peak around
$Q(\tau_3)$ though the initial excitation was ``flat'' up to the cut
off.

It will turn out in the next section that a large peak in the
amplification factor implies power--law evolution of the
characteristic wave number $Q(\tau)$ and thus of the tension
$\varphi(\tau)$. Whether the mode spectrum develops a pronounced peak as
in the above example or not, depends on the initial
conditions. Fig.~\ref{fig:bow-wave} suggests that a strong peak is
present at time $\tau_3$, because the dark grey area $\Delta
\epsilon_+=\Delta\epsilon_-$ is much larger then the light grey
area. In other words, the stored length will be strongly localized
around $Q(\tau)$ at time $\tau$, if the relaxed stored length
$|\Delta\epsilon_-|=\Delta\epsilon_+$ is much larger than the contour
length that was initially stored in the interval of width $\Delta Q$
around the wave number $Q$. To estimate the former, we first note,
that $A(q,\tau)$ decays exponentially to zero for $q>\sqrt{2}Q$, so
that we simply replace it by zero in Eq.~(\ref{eq:s-}). Then, because
many modes contribute to the remaining sum, we take the continuum
limit in mode space according to
\begin{equation}
  \label{eq:continuumlimit}
  \epsilon^0(q=q_n)\equiv\frac{\epsilon^0_n}{\pi/l} \;,
\end{equation}
so that we obtain
\begin{equation}
  \label{eq:epsilon_-int}
  \Delta\epsilon_-\approx -\mint{dq}{\sqrt{2}Q}{\infty}\epsilon^0(q) \;.
\end{equation}
The amount of contour length initially stored around $Q$ can be
estimated by the initial amplitude at $Q$, $\epsilon^0(Q)$, multiplied
by the width of the peak $\Delta Q$. Upon comparing $\epsilon^0(Q)
\Delta Q$ with $\Delta\epsilon_-$ we obtain the criterion
\begin{equation}
  \label{eq:alpha-estimate-integral}
  \mint{dq}{\sqrt{2}Q}{\infty}\epsilon^0(q) \gg \epsilon^0(Q) \Delta Q
 \;  \Leftrightarrow  \; \mbox{peak in $A(q,\tau)$}\;,
\end{equation}
to decide whether a peak is expected for a given value of
$Q(\tau)$. We refer to initial conditions as \emph{type~I} if they
guarantee that Eq.~(\ref{eq:alpha-estimate-integral}) holds after a
transient time that is much shorter than the overall relaxation time.

Assuming condition Eq.~(\ref{eq:alpha-estimate-integral}) to hold we
will show in the next section how \emph{approximate} power--law
relaxation of the dominant wave number $Q(\tau)$ and of the tension
$\varphi(\tau)$ emerges. In Sec.~\ref{sec:similarity_solutions} we
will see that \emph{exact} power--law solutions of
Eq.~(\ref{eq:force-cond-mode-space}) moreover arise from self--affine
initial conditions $\epsilon^0(q)\propto q^{-\beta}$. It will turn out
that a complete classification of all
generic~\footnotemark[\value{fnt}] relaxation scenarios in the bulk
can be given in terms of the roughness exponent $\beta$,
characterizing the initial contour undulations.

\subsection{Cascading of Stored Length (\emph{Type~I})}
\label{sec:cascade}
Provided that condition Eq.~(\ref{eq:alpha-estimate-integral}) is
fulfilled at a time $\tau$ larger than some suitable short transient
time, the relaxation has accumulated most of the stored length
$\epsilon$ in the peak around $Q(\tau)$. Undulations of wavelength
$Q^{-1}$ visibly dominate the rod contour. Furthermore, the sum
Eq.~(\ref{eq:s+}) representing $\Delta \epsilon_+$ is dominated by the
modes around $Q(\tau)$, which simplifies its evaluation significantly.
Yet, one still has to discriminate two limiting cases.

\emph{Intermediate Asymptotics:} The peak of the amplification factor
covers many modes, i.e., the width $\Delta Q$ of the amplification
peak, as defined in Eq.~(\ref{eq:width}), is much larger than the mode
spacing $\pi/l$, or
\begin{equation}
  \label{eq:spas}
  Q l\gg 2\pi\sqrt{\alpha} \;.
\end{equation}
Then many modes contribute to both, $\Delta \epsilon_-$ and $\Delta
\epsilon_+$ and the corresponding sums Eqs.~(\ref{eq:s-},~\ref{eq:s+}) can be
converted into integrals, as has already been done in
Eq.~(\ref{eq:epsilon_-int}) for $\Delta \epsilon_-$ to obtain the
criterion Eq.~(\ref{eq:alpha-estimate-integral}). The continuum limit
for $\Delta \epsilon_+$ reads
\begin{equation}
  \label{eq:s+-int} \Delta \epsilon_+\approx \mint{dq}{0}{ \sqrt{2}Q}
  \epsilon^0(q) \left\{\exp\left[2\tau
  q^2\left(2Q^2-q^2\right)\right]-1\right\} \;.
\end{equation}
Since by assumption, Eq.~(\ref{eq:alpha-estimate-integral}), the
integrand in Eq.~(\ref{eq:s+-int}) has a pronounced maximum, it can be
evaluated by a saddle point approximation, replacing it effectively
by the area $\Delta Q \exp(2 \alpha)$ under the amplification peak
$A(q)$ multiplied by $\epsilon^0(Q)$.
\begin{equation}
  \label{eq:epsilon_+approx}
  \Delta \epsilon_+\approx\epsilon^0(Q) \Delta Q \exp\left(2 \alpha\right) \;.
\end{equation}
The conservation of the stored length, Eq.~(\ref{eq:s+=s-}), implies
\begin{equation}
  \label{eq:I1=I2}
  0=\Delta \epsilon(\tau)\approx\epsilon^0(Q) \Delta Q \exp\left(2\alpha\right)-\mint{dq}{\sqrt{2}Q }{\infty}
  \epsilon^0(q) \;.
\end{equation}
Since the first term on the right--hand--side of Eq.~(\ref{eq:I1=I2})
depends exponentially on the parameter $\alpha$, the latter is slaved
to be time--independent up to logarithmic corrections,
\begin{equation}
  \label{eq:alpha-const}
  \alpha= \mbox{constant} +  O(\ln\tau) \;.
\end{equation}
Recalling the definition of $\alpha$, Eqs.~(\ref{eq:alpha-def}), and
using Eq.~(\ref{eq:phifromQ}), one finds for the tension
\begin{equation}
  \label{eq:scaling-Q-phi}
  \varphi(\tau)= Q(\tau)^2\propto \tau^{-1/2} \;,
\end{equation}
which proves Eqs.~(\ref{eq:ft},\ref{eq:qt}) for the intermediate
asymptotics of \emph{type~I} up to logarithmic corrections. While the
peak position is thus migrating to lower wave numbers according to the
power law $Q(\tau)\propto\tau^{-1/4}$, its width shrinks accordingly,
$\Delta Q\propto \tau^{-1/4}$. Consequently the number of discrete
modes under the amplification peak decreases.

\emph{Ultimate Staircase Relaxation:} When $\Delta Q$ eventually becomes
smaller than the mode spacing $\pi/l$, the contour of the rod starts
to be dominated by the discrete wavenumber $q_{n^\star}$ closest to
$Q(\tau)$. Thus it is no longer legitimate to approximate $\Delta
\epsilon_+$ by an integral. On the contrary, in the limit
\begin{equation}
  \label{eq:sma}
  Q l \ll 2\pi\sqrt{\alpha}
\end{equation}
the sum in Eq.~(\ref{eq:s+}) should be replaced by the single dominant
element corresponding to the index $n^\star$,
\begin{equation}
  \label{eq:spa-s+}
  \Delta \epsilon_+=\epsilon^0_{n^\star}\exp\left[2 \tau q_{n^\star}
  (2Q^2-q_{n^\star}^2)\right] \;.
\end{equation}
In contrast, the sum $\Delta \epsilon_-$ representing the destroyed stored
length has contributions from many modes even in the limit
Eq.~(\ref{eq:sma}) and it can still be approximated by the integral
Eq.~(\ref{eq:epsilon_-int}). The parity of created and destroyed
stored length,
Eq.~(\ref{eq:s+=s-}), now takes the form
\begin{equation}
  \label{eq:spa}
  \epsilon^0_{n^\star}\exp\left[2 \tau q_{n^\star}
  (2Q^2-q_{n^\star}^2)\right] \approx \mint{dq}{\sqrt{2}Q }{\infty}
  \epsilon^0(q) \epsilon^0_n \;.
\end{equation}
As below Eq.~(\ref{eq:I1=I2}) we conclude that the exponent on the
left-hand-side has to stay constant in time up to logarithmic
contributions. By using the definition of $Q$,
Eq.~(\ref{eq:characteristic-wnumber}), this implies, that the tension
is equal to the Euler force corresponding to the mode $n^{\star}$,
\begin{equation}
  \label{eq:spa-2}
  \varphi\approx q_{n^\star}^2 \;,
\end{equation}
as long as $n^{\star}$ is indeed the dominant mode. In fact, the
discrete $n^{\star}$ is a time--dependent quantity that evolves in
steps and approaches $1$ in the final stage of the relaxation, which
corresponds to the first Euler buckling mode.

\begin{figure}
       \psfrag{x}{$\log\left(\tau/\tau_0\right)$}
       \psfrag{y}{$\log\left(\varphi/\varphi_1\right)$}
       \psfrag{-18.4}{$\log(\tau_{100}/\tau_0)$}
       \psfrag{-10}{$-10$}
       \psfrag{0}{$0$}
       \psfrag{4}{$4$}
       \psfrag{8}{$8$}
       \psfrag{N=1h20}{$N=10^{20}$}
       \psfrag{N=1h6}{$N=10^6$}
       \psfrag{N=100}{$N=100$}
       \psfrag{1/2}{$1/2$}
  \begin{center}
         \includegraphics[width=\columnwidth]{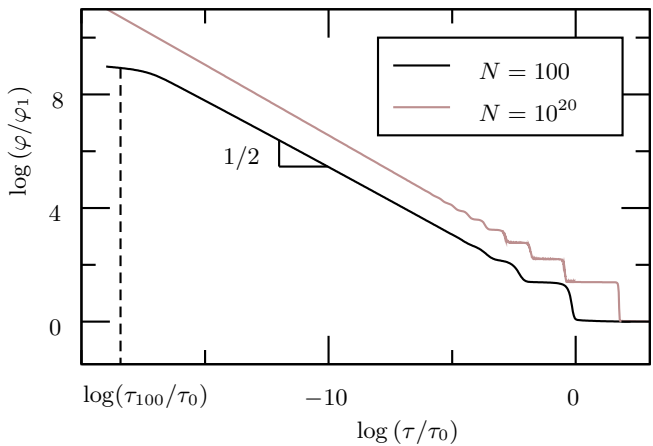}
\caption{Typical tension relaxation of a rod with initial conditions
  satisfying Eq.~(\ref{eq:alpha-estimate-integral}) from a numerical
  solution of Eq.~(\ref{eq:force-cond-mode-space}). As in the example
  of Fig.~\ref{fig:bow-wave} we chose the particular initial condition
  $\epsilon^0_{n\leq N}=\mbox{const.}$ and $\epsilon^0_{n>N}=0$. For two
  values of $N$ the graph displays the tension $\varphi$ versus time
  $\tau$ in units of critical force $\varphi_1=(\pi/l)^2$ and the
  typical relaxation time $\tau_0=l^4$, respectively. In the case
  $N=100$ it is seen that the intermediate asymptotic power law
  $\varphi\propto \tau^{1/2}$ is valid in the time window $\tau_N
  \ll\tau\ll\tau_0$, where $\tau_N\equiv N^{-4}\tau_0$ is the
  relaxation time of the highest excited mode. The extreme case
  $N=10^{20}$ illustrates the asymptotic behavior of the staircase
  regime for large $N$.}  \label{fig:force-time}
\end{center}
\end{figure}

To illustrate the above discussion, Fig.~\ref{fig:force-time} displays
the (normalized) line tension $\varphi(\tau)$ obtained from the
numerical solution of the implicit
Eq.~(\ref{eq:force-cond-mode-space}) for $Q(\tau)$. The shown
relaxation scenario is characteristic of the dynamics for the class of
initial conditions satisfying the condition in
Eq.~(\ref{eq:alpha-estimate-integral}). For short times, one observes
after a short transient period a smooth intermediate asymptotic
power--law behavior $\varphi(\tau)\sim \tau^{-1/2}$, which for long
times develops staircase--like oscillations with plateaus at
$\varphi_n=n^2 \varphi_1$, in agreement with the above derivation.

\subsection{Exact Similarity Solutions (\emph{Type II})}
\label{sec:similarity_solutions}

\begin{figure}
  \psfrag{x}{$\beta$} \psfrag{yyy}{$\sqrt{\alpha_\beta}$}
  \psfrag{I}{\hspace*{-.1cm}(II)} \psfrag{II}{\hspace{-.05cm}(I)}
  \psfrag{III}{} \psfrag{0}{$0$} \psfrag{1}{$1$} \psfrag{3}{$3$}
  \psfrag{+inf}{$+\infty$} \psfrag{2}{$2$} \psfrag{-inf}{$-\infty$}
  \psfrag{pathological}{\hspace{-0cm}ground state} 
  \psfrag{exact}{\hspace{-0cm}$\hspace{.3cm}\stackrel{\displaystyle
      \text{exact similarity}}{\text{solutions}}$}
  \psfrag{peaked solutions}{\hspace{-0cm}$\stackrel{\displaystyle
      \text{localization}}{\text{in mode space}}$}
  \begin{center}
    \includegraphics[width=\columnwidth]{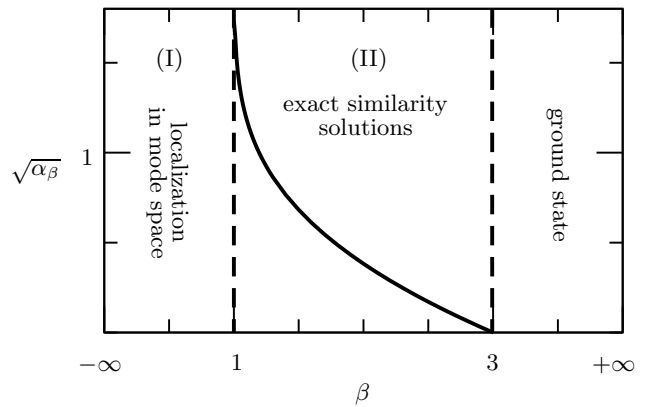}
\caption{The amplitude $\sqrt{\alpha_\beta}$ in the power law
  Eq.~(\ref{eq:ansatz}) for the line tension $\varphi$ in the bulk of
  a relaxing rod with power--law initial conditions
  $\epsilon^0(q)\propto q^{-\beta}$. For $\beta<1$ an upper cutoff is
  required to keep the stored length finite. The relaxation then
  proceeds via the cascading of a localized peak in mode space
  (\emph{type I} behavior), as depicted in
  Fig.~\ref{fig:greens-function}(a) and explained in
  Sec.~\ref{sec:cascade}. For $\beta>3$, the rod is essentially in the
  ground state from the beginning.  The interval $1<\beta<3$ of
  \emph{type~II} initial conditions comprises the exact similarity
  solutions derived in Sec.~\ref{sec:similarity_solutions}.  Thermal
  initial conditions correspond to $\alpha_{\beta=2}\approx 0.146$.}
\label{fig:critical-alphas} \end{center}
\end{figure}

Besides the cascading of stored length that is strongly localized in
mode space, there is a different mechanism giving rise to the power
law Eq.~(\ref{eq:ft}). This is revealed by explicitely searching for
similarity solutions of Eq.~(\ref{eq:force-cond-mode-space}) under the
condition that many modes contribute to the relaxation dynamics and
the sums in Eq.~(\ref{eq:force-cond-mode-space}) can again be
converted into integrals. In contrast to the intermediate asymptotic
power--law solutions obtained in Sec.~\ref{sec:cascade}, which obeyed
$\alpha=$ const.\ only up to logarithmic corrections, solutions
$Q(\tau)$ of the continuum limit of the constraint
Eq.~(\ref{eq:force-cond-mode-space}),
\begin{equation}
  \label{eq:constr-continuum}
  0=\mint{dq}{l^{-1}}{\infty}\epsilon^0
\left(q\right)\left\{\exp[2 \tau q^2 (2 Q^2- q^2)]-1\right\}\;,
\end{equation}
can be found that \emph{exactly} obey
\begin{equation}
  \label{eq:alfa-const}
  \alpha\equiv Q^4\tau\stackrel{!}{=}\mbox{const.} 
\end{equation}
Inserting the ansatz
\begin{equation}\label{eq:ansatz}
\varphi(\tau)= Q^2(\tau) =   (\alpha/\tau)^{1/2}
\end{equation}
with a yet undetermined time--independent parameter $\alpha$ into
Eq.~(\ref{eq:constr-continuum}) and changing variables $q\rightarrow q
(\alpha/\tau)^{1/4}$, we obtain
\begin{equation}
  \label{eq:nondimforce}
  0=\mint{dq}{\tau^{1/4}/(\alpha l)}{\infty}\epsilon^0
\left[q \left(\alpha/\tau\right)^{1/4}\right]\left\{\exp[2\alpha(-q^4 +2 q^2)]-1\right\}\;.
\end{equation}
This is mathematically equivalent to
Eq.~(\ref{eq:force-cond-mode-space}) as long as the integral is not
sensitive to its (small) lower bound, so that the latter can
effectively be taken to be zero. Then, for the integral to be
independent of time, the initial condition has to be of power--law
form,
\begin{equation}
  \label{eq:self-affine-inicond}
  \epsilon^0(q)=\Lambda^{1-\beta} q^{-\beta} \;,
\end{equation}
where the length $\Lambda$ has to be introduced on dimensional
grounds. The numerical solutions $\alpha_\beta$ of
Eq.~(\ref{eq:nondimforce}) are depicted in
Fig.~\ref{fig:critical-alphas} as a function of $\beta$. As can be
seen, the roughness exponent $\beta$ is not completely arbitrary. In
fact, no finite solutions for $\alpha$ exist outside the interval
$1<\beta<3$. We refer to Eq.~(\ref{eq:self-affine-inicond}) with
$1<\beta<3$ as \emph{type~II} initial conditions.

For \emph{type~II} initial conditions the ansatz
Eq.~(\ref{eq:self-affine-inicond}) solves Eq.~(\ref{eq:nondimforce})
exactly in the limit that the lower bound tends to zero, or for times
\begin{equation}
  \label{eq:shorttimes}
  \tau\ll (\alpha l)^4\;.
\end{equation}
Then the initial conditions Eq.~(\ref{eq:self-affine-inicond})
parameterize a novel class of power--law solutions (to our knowledge)
not seen previously. The algebraic decay law can in this case be
attributed to the self--affine geometry of the initial conformation.
Note that the weakly bending condition expressed in
Eqs.~(\ref{eq:wkb},~\ref{eq:epsilon}) requires
\begin{equation}\label{eq:wbLambda}
 \epsilon^0\approx
 \mint{dq}{l^{-1}}{\infty}{\epsilon^0(q)} = 
 \frac{\left(l/\Lambda\right)^{\beta-1}}{\beta-1} \ll 1\;,
\end{equation}
i.e.\ $\Lambda\gg l$ for rod sections of length $l$.  An important
example for these initial conditions is provided by the contour of a
stiff polymer in thermal equilibrium ($\beta=2$)
\cite{doi-edwards:86,yamakawa}, for which the length $2\Lambda/ \pi$
is identified as the persistence length of the polymer, which indeed
has to be much larger then the length of the polymer in the weakly
bending limit. The corresponding amplification factor $A(q)$ with
$\alpha_{\beta=2}\approx 0.146$ as a function of $q$ is shown in
Fig.~\ref{fig:greens-function}(b).

In contrast to the \emph{type~I} intermediate asymptotics discussed in
Section~\ref{sec:cascade}, the \emph{type~II} dynamics for $1<\beta<3$
is not necessarily governed by a characteristic wave number that
visibly dominates the contour undulations.  For
$\alpha_\beta\lesssim1$ the amplification factor rather acts as a
time--dependent low pass filter cutting off the mode amplitudes with
wave numbers larger than $\sqrt{2}Q(\tau)$. This is to be contrasted
with the situation in Fig.~\ref{fig:greens-function}(a), where the
amplification factor is strongly peaked around $Q(\tau)$. Only in the
limit $\beta\to1$, do the self--affine initial conditions
Eq.~(\ref{eq:self-affine-inicond}) satisfy the condition
Eq.~(\ref{eq:alpha-estimate-integral}) that guarantees a large peak in
the amplification factor, thus giving way to the scenario described in
Sec.~\ref{sec:cascade}, but \emph{without} logarithmic corrections.
Because of the latter, we consider the initial conditions
Eq.~(\ref{eq:self-affine-inicond}) with $\beta\gtrsim 1$ as
sufficiently distinct from the above defined general \emph{type~I}
solutions to justify the classification under \emph{type~II} even for
$\beta\to1$.

The fact that the value $\alpha=\alpha_\beta$ that solves
Eq.~(\ref{eq:nondimforce}) diverges as $\beta$ approaches $1$ from
above indicates an unphysical situation. The initially stored length
in modes with large wave numbers grows without bound and the integral
over $\epsilon^0(q)$ diverges. Hence for $\beta\leq 1$ power--law initial
conditions as in Eq.~(\ref{eq:self-affine-inicond}) are only
well--defined with an upper cut--off, say the wave number $q_N$
corresponding to the highest excited mode with index $N$.
Furthermore, with Eq.~(\ref{eq:self-affine-inicond}) the weakly
bending condition now requires $q_N\Lambda\ll 1$. Then, for times
$\tau$ such that $Q(\tau)\ll q_N$ the initial conditions automatically
fulfill the criterion Eq.~(\ref{eq:alpha-estimate-integral}) for the
\emph{type~I} scenario developed in Section~\ref{sec:cascade}. In this
sense, the exponents $\beta \leq 1$ are representative of \emph{type~I}
relaxation scenarios that are characterized by a localization in mode
space. 

Finally, upon expanding the integrand of Eq.~(\ref{eq:nondimforce})
into a Taylor series for small $q$ it is seen that for $\beta >3$ the
integral would be dominated by the lower bound, indicating the
breakdown of the continuum approximation.  The sum in
Eq.~(\ref{eq:force-cond-mode-space}) is then dominated by its first
term, the first Euler buckling mode. This yields a tension of about
$\varphi\approx\varphi_1\propto l^{-2}$: A confined buckled rod with
the initial condition Eq.~(\ref{eq:self-affine-inicond}) and $\beta>3$
is essentially in the ground state from the beginning.

We have thus achieved a complete classification of the possible
relaxation scenarios for all generic~\footnotemark[\value{fnt}]
initial conditions for the key--problem of a longitudinally confined
rod, which was previously studied in numerical simulations
\cite{golubovic-moldovan-peredera:98}. The results are summarized in
Fig.~\ref{fig:critical-alphas}. The power--law decay Eq.~(\ref{eq:ft})
of the tension emerges as a quite universal feature of the problem,
whereas for the accompanying conformational relaxation one has to
distinguish two fundamentally different scenarios classified as
\emph{type~I} and \emph{type~II} according to two corresponding types
of initial conditions.

\subsection{Conformational Relaxation}

\begin{figure}
  \psfrag{e}{\hspace{-.5cm}$e(q,\tau)$} \psfrag{Q}{$Q(\tau )$}
  \psfrag{uc}{$\sqrt{2}Q(\tau)$} \psfrag{0}{$0$}
  \psfrag{q}{\hspace{3.3cm}$q$} \psfrag{0}{$0$} \psfrag{lc}{$L^{-1}$}
  \psfrag{b2}{\hspace{-.2cm}$\beta=2$}
  \psfrag{b0}{\hspace{-.2cm}$\beta=0$, $\alpha\gg1$}
    \includegraphics[width=\columnwidth]{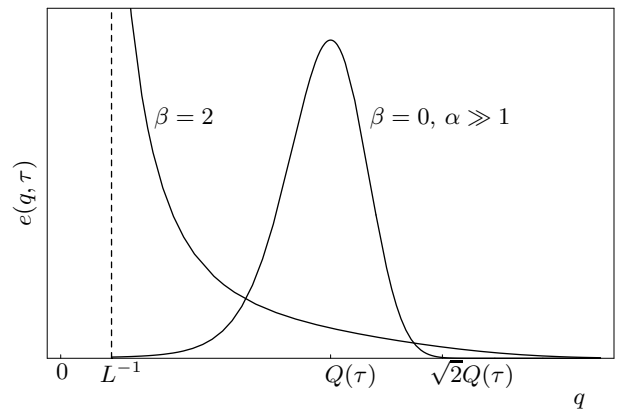}
    
\caption{The situation in mode space after time $\tau$ for
representative initial conditions of \emph{type~I} ($\beta=0$,
$\alpha\gg1$, cutoff $q_N\gg Q$ as in Fig.~\ref{fig:bow-wave}) and
\emph{type~II} (thermal initial conditions $\beta=2$) with the same
total stored length $\epsilon^0$.}
\label{fig:terminal-relaxation} 
\end{figure}

For \emph{type~I} initial conditions the intermediate asymptotic
dynamics is completely governed by the characteristic wavelength
$Q^{-1}$. The latter directly determines the tension and visibly
dominates the real--space image of the contour, so that tension decay
and conformational relaxation occur hand in hand. A markedly different
scenario results for \emph{type~II} initial conditions. To appreciate
the difference, consider the representative distributions of stored
length in mode space depicted in Fig.~\ref{fig:terminal-relaxation}.
At time $\tau$, the stored length that was initially distributed in
the tails $q\geq \sqrt2 Q(\tau)$ has been accumulated around
$Q(\tau)$. Due to the substantially different relative weight of these
tails in the initial conditions, the corresponding distributions at
time $\tau$ look utterly different.  While over a time interval
$16\tau$, the undulations dominating the real space image of the
contour will have doubled their wavelength under \emph{type~I}
conditions, the corresponding evolution of $Q(\tau)$ will have hardly
any noticeable consequences on the real space image of a
\emph{type~II} contour, which is dominated by undulations of much
longer wavelengths that are practically stationary on this time scale.

As we pointed out in the Introduction, the relaxation of a laterally
confined rod that was discussed in the present section, can also be
considered an idealization of the situation in the bulk of a long
stiff rod with \emph{free} ends that was initially under high
pressure, which also has been simulated \cite{spakowitz:2001}. At the
free ends, the tension $\varphi(s,\tau)$ has to vanish as a
consequence of the boundary conditions Eq.~(\ref{eq:free-bc}). In the
following we face the question how it falls off between the bulk and
the ends.

\section{The Relaxation for Open Boundaries}
\label{sec:free-relaxation}
In the present section and in the appendix, we develop a method to
treat situations where the tension exhibits substantial spatial
variations along the rod. The basic idea is as follows. The major
variation of the tension, namely from its bulk value to zero at the
open boundaries, occurs within a (time--dependent) boundary layer of a
yet unknown length $\lambda(\tau)$. In the next section we will
motivate the crucial length scale separation between $\lambda(\tau)$
and $Q^{-1}(\tau)$.  It will allow us to apply our leading order
results from Sec.~\ref{sec:lin} \emph{locally} on an intermediate
scale $l(\tau)$ ($Q^{-1}\ll l \ll\lambda$), over which the tension
does not change appreciably. This in turn will enable us to derive
closed equations for a suitably coarse--grained tension profile
$\varphi_l(s,\tau)$ in Sec.~\ref{sec:adiabatic_approximation}. This
adiabatic approximation will eventually be justified by a consistency
check. (Its precise relation to the regular perturbation scheme of
Sec.~\ref{sec:lin} will be clarified at the end of
Sec.~\ref{sec:free-relaxation} and in the appendix.) Our discussion of
the boundary layer problem will parallel the discussion in
Sec.~\ref{sec:lin} in discerning again \emph{type~I} and
\emph{type~II} initial conditions. Thereby we will in particular
recover for the bulk our earlier results, which were based on the
assumption of longitudinal confinement.

\subsection{Length Scale Separation}
\label{sec:adi}
Technically, to address spatial variations in the tension profile,
which could be discarded as of higher order in $\epsilon$ in the
regular perturbation scheme of Sec.~\ref{sec:leading_order}, we need
to push the analysis beyond the leading order.  In
Sec.~\ref{sec:ampelfaktor} we have determined the time evolution of
the transverse displacements $\vec r_\perp(s,\tau)$ of a rod section
of length $l$ to leading order in $\epsilon$. By inserting with the
help of Eq.~(\ref{eq:local-const}) the result back into the
(higher--order) equation of motion Eq.~(\ref{eq:parallel-eq-motion})
for $\vec r_\|$, we can iteratively estimate the order of magnitude of
the spatial variation of the tension, which is determined by the
nonlinear terms.

\begin{figure}
      \psfrag{L}{$\lambda$}
      \psfrag{l}{$l$}
      \psfrag{f(s,t)}{$\varphi(s)/\varphi_\infty$}
      \psfrag{e(s,t)}{$\epsilon(s)$}
      \psfrag{fl}{$\frac{\varphi_l(s)}{\varphi_\infty}$}
      \psfrag{el}{$\epsilon_l(s)$}
      \psfrag{s}{$s$}
      \psfrag{Q}{$Q_\infty^{-1}$}
 \begin{center}
        \includegraphics[width=\columnwidth]{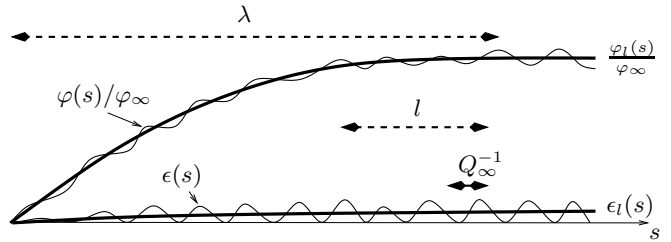}
\caption{The length $\lambda$ over which the tension increases towards the bulk
  value is much larger than the characteristic length $Q_\infty^{-1}$
  of the transverse undulations in the bulk, as expressed by the
  inequality in Eq.~(\ref{eq:length-scale-sep}). The slowly varying part
  $\varphi_l(s)$ of the tension is obtained upon averaging over the
  coarse--graining scale $l$, satisfying the condition in
  Eq.~(\ref{eq:adiabatic_length}).}
   \label{fig:coarse-graining}
 \end{center}
\end{figure}

Note that the leading--order solution for $\vec r_\perp(s,\tau)$
depends on the force history of the particular rod section under
consideration, which enters via the characteristic wave number
$Q(\tau)$. We recall from our discussion of the amplification factor
in Sec.~\ref{sec:ampelfaktor} that $\sqrt{2}Q(\tau)$ acts as an
effective ultra--violet cutoff for the contour undulations. From
Eqs.~(\ref{eq:local-const},~\ref{eq:epsilon}), we thus have for
example
\begin{equation}\label{eq:newpc}
 2r_\|'''' = (\vec r_\perp'^2)''' \leq  Q^3\,O(\epsilon) \;.
\end{equation}
Time derivatives are estimated by recourse to
Eq.~(\ref{eq:perp-eq-motion}). Applying this reasoning to
Eq.~(\ref{eq:parallel-eq-motion}) after differentiating with respect
to arc length $s$, one eventually finds
\begin{equation}
  \label{eq:tension-variations}
  \varphi''\leq Q^4 \, O(\epsilon) \;
\end{equation}
for the order of magnitude of the tension variations. Generalizing
$Q(\tau)\to Q(s,\tau)$ to allow for a slow spatial variation of the
characteristic wavelength, we can integrate
Eq.~(\ref{eq:tension-variations}) from one end of the contour, where
$\varphi=0$ and $Q=0$, towards the bulk, where $\varphi\equiv
\varphi_\infty = Q^2 \equiv Q_\infty^2$. (Here and in the following,
we write symbolically ``$\infty$'' to refer to regions deep in the
bulk.)  Since $Q\leq Q_\infty$, we can infer
\begin{equation}
  \label{eq:lambda-low}
  Q_\infty^2=\varphi_\infty=\mint{ds}{0}{\lambda}\mint{d\hat
    s}{0}{s}\varphi''\leq Q_\infty^4 \lambda^2\; O(\epsilon ) \nonumber \;,
\end{equation}
by integrating through the boundary layer of length
$\lambda(\tau)$. From this we read off a lower bound for the order of
magnitude of $\lambda$,
\begin{equation}
  \label{eq:length-scale-sep}
  \Rightarrow \quad (\lambda Q_\infty)^{-1}\leq
  O\left(\sqrt{\epsilon}\right) \;.
\end{equation}
For small $\epsilon\to0$, we thus have a strong length scale
separation between wavelength $Q_\infty^{-1}$ of the dynamically most
active contour undulations and the scale $\lambda$ of the substantial
tension variations, i.e.\ $\lambda\gg Q_\infty^{-1}$. It allows us to
define a length $l(t)$ intermediate between the characteristic scales
$Q_\infty^{-1}(t)$ and $\lambda(t)$, so that
\begin{equation}
  \label{eq:adiabatic_length}
  1\ll Q_\infty l\ll \epsilon^{-1/2}\;.
\end{equation} 
Fig.~\ref{fig:coarse-graining} illustrates the relation between the
various lengths. An immediate consequence of the inequalities
Eqs.~(\ref{eq:length-scale-sep},~\ref{eq:adiabatic_length}) is that we
can imagine the free rod at any time as consisting of rod sections of
length $l\gg Q_\infty^{-1}$, each of which is subject to a uniform
``average'' tension. After specifying this average we will be ready to
\emph{locally} apply our results of Sec.~\ref{sec:lin} to the problem
of a rod with free ends in the next sections.

\subsection{Adiabatic Approximation}
\label{sec:adiabatic_approximation}

The length scale separation observed in the previous section suggests
to look for a mathematical description of the ``substantial''
variation of the tension on the scale $\lambda$ without its
complicated wiggling on the ``micro--scale'' $Q_\infty^{-1}$, which is
at most of order $Q_\infty^2 O(\epsilon)$. The natural way to get rid
of the short wavelength fluctuations without loosing the substantial
part is to consider coarse--grained quantities that are averaged over
the intermediate scale $l$. More precisely, we define for any
arc--length dependent quantity $g(s)$ a corresponding coarse--grained
quantity $g_l(s)$ by
\begin{equation}
  \label{eq:cg-average} g_l(s)\equiv
  \frac{1}{l}\mint{d\sigma}{-l/2}{l/2}g(s+\sigma) \;.
\end{equation}
It will turn out that for the quantities of interest, this average is
actually \emph{independent} of $l$ to leading order in $\epsilon$, if
$l$ obeys the double inequality Eq.~(\ref{eq:adiabatic_length}). For
the tension $\varphi(s,\tau)$ this was already established in
Sec.~\ref{sec:adi}.

A closed equation for the coarse--grained tension $\varphi_l$ can now
be derived from the full equations of motion, Eqs.~(\ref{eq:EOM}).
Upon integrating the longitudinal Eq.~(\ref{eq:parallel-eq-motion})
with respect to the arclength and using the free boundary conditions
Eq.~(\ref{eq:free-bc}), we at first obtain an explicit equation for
the spatially varying tension $\varphi(s,\tau)$ before
coarse--graining
\begin{eqnarray}
  \label{eq:tension-integral-eq}
  \varphi(s,\tau)&=&\hat{\zeta}\mint{d\tilde s}{0}{s}\partial_\tau
  r_\parallel-(1-\hat{\zeta})\mint{d\tilde s}{0}{s}\vec r_\perp'\partial_\tau
  \vec r_\perp+ \nonumber \\
  & &+r_\parallel'''(s)+\varphi(s,\tau) r_\parallel'(s) \;.
\end{eqnarray}
Here $\hat \zeta=\zeta_\parallel/\zeta_\perp= 1/2$ is the ratio
between the transverse and longitudinal friction coefficients. Using
our knowledge about the bulk we now show that only the first term on
the right hand side is able to produce a term of the order of the
tension $\varphi_\infty$ in the bulk. Counting arc--length derivatives
in orders of $Q_\infty$ in the spirit of Sec.~\ref{sec:adi},
$\varphi_\infty$ is estimated as of order $O(Q_\infty^2)$. The crucial
fact that $\sqrt{2}Q_\infty$ acts as a high--wave--number cutoff for
the contour fluctuations together with the local constraint
Eq.~(\ref{eq:local-const}) implies that the last two terms are
$O(\epsilon Q_\infty^2)$, thus always small compared to the bulk
tension $\varphi_\infty$. The same reasoning can be applied to the
second term on the right hand side after suitable partial
integrations,
\begin{eqnarray}
  \label{eq:anisotropy-term}
  \left|\mint{d\tilde{s}}{0}{s}\vec r_\perp' \partial_\tau \vec
  r_\perp\right|
  &\stackrel{\mbox{Eq.~(\ref{eq:perp-eq-motion})}}{=}&\left|\mint{d\tilde{s}}{0}{s}\vec
  r_\perp' (-\vec r_\perp''''-(\varphi\vec r_\perp')')\right|
  \nonumber \\ &\stackrel{\varphi<\varphi_\infty}{\leq}&|\vec r_\perp' \vec
  r_\perp'''|+\vec r_\perp''^2+\varphi_\infty \vec r_\perp'^2 \nonumber \\
  &=&O(\epsilon Q_\infty^2) \nonumber \;.
\end{eqnarray}
Consequently the necessary $O(Q_\infty^2)$ term on the right hand side
of Eq.~(\ref{eq:tension-integral-eq}) must be the one depending on the
longitudinal velocity $\partial_\tau r_\parallel$. It represents the
pressure that is generated in the rod by the outward motion of the
relaxing boundary layer. Differentiating
Eq.~(\ref{eq:tension-integral-eq}) twice with respect to arc length
and integrating over time we can therefore write to leading order
\begin{equation}
  \label{eq:leading-order-tension-eq}
  \hat \zeta^{-1}\Phi''=\hat \zeta^{-1}\mint{d\hat \tau}{0}{\tau}\varphi''(s,\hat\tau)=
  \epsilon(s,\tau)-\epsilon(s,0) \;.
\end{equation}
Since we are interested in the long wavelength fluctuations of
$\varphi$, we average this equation over the length $l$ and obtain
\begin{equation}
  \label{eq:coarse-graining}
  \hat \zeta^{-1}\Phi''_l(s,\tau) = \Delta \epsilon_l(s,\tau)
  \;.
\end{equation}
The physical interpretation of this important result is that the
release $\Delta \epsilon_l \equiv
\epsilon_l(s,\tau)-\epsilon_l(s,0)<0$ of stored length corresponds to
a negative curvature in the time integrated tension profile,
$\Phi_l''<0$. Stored--length release acts as a source for spatial
variations of the time integrated tension.  For increasing arclength
$s$ we expect the tension to saturate, $\Phi''(s\to \infty)\to 0$,
corresponding to a conserved stored length in the bulk, $\Delta
\epsilon_l(s\to \infty)=0$, just as we argued throughout
Sec.~\ref{sec:lin}. (Here and in the following we write symbolically
``$\infty$'' for contour elements deep in the bulk.) In this sense,
Eq.~(\ref{eq:coarse-graining}) generalizes the conservation law
Eq.~(\ref{eq:global-constraint}) for the bulk to the boundary layer.
For the interested reader a second, more formal derivation of
Eq.~(\ref{eq:coarse-graining}) via the method of
homogenization~\cite{holmes:95} is given in Appendix
\ref{sec:homogenization}.

In order to close Eq.~(\ref{eq:coarse-graining}) we need an expression
for the stored length release $\Delta \epsilon_l$ on the scale $l$ as
a function of $\Phi_l$. Here, we can simply refer back to
Sec.~\ref{sec:ampelfaktor}. There we have dealt with one
``coarse--graining element'' of length $l$ to leading order in
$\epsilon$. We recall that a crucial ingredient of the (regular)
perturbation calculation in Sec.~\ref{sec:ampelfaktor} was that we
could neglect spatial variations of $\varphi(s,\tau)$ to leading order
in $\epsilon$. The length scale separation observed in
Sec.~\ref{sec:adi} shows that we can neglect them on the scale $l$,
which is much larger than the characteristic wavelength $Q^{-1}$ of
the dynamically most active transverse modes. Therefore, our above
perturbative results can be used in the present boundary layer
calculations. With the
Eqs.~(\ref{eq:greens_fct}--\ref{eq:ampl-scalingform1}) we can describe
the evolution of $\epsilon_l(s,\tau)$ as a function of
$\varphi_l(s,\tau)$ by identifying the coarse--grained quantities with
the corresponding spatial averages in
Eqs.~(\ref{eq:wkb-condition},~\ref{eq:spatial-force-average}). This
identification constitutes the adiabatic approximation.

The stored length release in the continuum limit is now taken over
from the right hand side of Eq.~(\ref{eq:constr-continuum}),
\begin{equation}
  \label{eq:locally-stored-length} \Delta
  \epsilon_l(s,\tau)=\mint{dq}{l^{-1}}{\infty}\epsilon^0
  \left(q\right)\left[e^{2 \tau q^2 \left[2 Q_l(s,\tau)^2-
  q^2\right]}-1\right] \;
\end{equation}
with $Q_l(s,\tau)$ the spatially weakly varying, adiabatic quantity.
There are two things to remark about equation
(\ref{eq:locally-stored-length}). First, the use of an integral
instead of a sum is legitimate, if the integral is not dominated by
its lower bound. We infer from the length scale separation
Eq.~(\ref{eq:adiabatic_length}) that this is the case if the integrand
is dominated by wave numbers close to the effective upper cutoff
$\sqrt{2} Q_l$. The wiggles on scale $Q^{-1}$ are then the major
source for the release of stored length.  Second, in the most general
case, we should allow for a weak spatial dependence not only of $Q_l$
but as well of $\epsilon^0(q)$ by writing $\epsilon^0(q,s)$. For
simplicity, we neglect such a spatial dependence in the initial
conditions and focus on statistically {\em uniform} initial
excitations.

Upon inserting Eq.~(\ref{eq:locally-stored-length}) into
Eq.~(\ref{eq:coarse-graining}) and using $Q_l^2(s,\tau)\equiv
\Phi_l(s,\tau)/2 \tau$ we obtain a closed differential equation for
$\Phi_l$:
\begin{equation}
  \label{eq:classical-mech-ode-for-Phi}
  \Phi_l''(s,\tau)=\hat \zeta \mint{dq}{l^{-1}}{\infty} \epsilon^0(q)
  \left[e^{2  q^2\left[\Phi_l(s,\tau)-q^2\tau\right]}-1\right] \; .
\end{equation}
Specializing to the left end of a semi--infinite rod, this equation
has to be solved for the boundary conditions $\Phi_l=0$ (no force) at
the end and $\Phi_l''=0$ at $s\rightarrow \infty$ (conserved stored
length in the bulk).

The differential equation (\ref{eq:classical-mech-ode-for-Phi}) is of
a type frequently encountered in classical mechanics: By interpreting
$\Phi_l$ as the position of a particle (mass $\equiv 1$) and $s$ as the
time variable, Eq.~(\ref{eq:classical-mech-ode-for-Phi}) represents
Newton's equation,
\begin{equation}
  \label{eq:newton}
  \Phi_l''(s)=-\partial_{\Phi_l} U(\Phi_l) \; ,
\end{equation}
for a particle moving in a potential $U(\Phi_l)$,
\begin{equation}
  \label{eq:potential}
  U(\Phi_l)=\hat \zeta \mint{dq}{l^{-1}}{\infty} \epsilon^0(q)  
  \left[\Phi_l-\frac{1-e^{2q^2\Phi_l}}{2 q^2}e^{-2  q^4 \tau}\right] \;. 
\end{equation}
For fixed time $\tau$, the negative of this potential is $U-$shaped as
a function of $\Phi_l$. The mechanical analogon to our task is to find
the instanton solution $\phi_l$, that approaches the location of the
maximum of $U(\phi_l)$ as $s\to\infty$. Eq.~(\ref{eq:newton}) can be
integrated numerically for all times and arbitrary initial conditions
$\epsilon^0(q)$. Having obtained $\Phi_l(s,\tau)$, the tension profile
$\varphi(s,\tau)$ is extracted by taking the derivative,
\begin{equation}
  \label{eq:extract-tension1}
  \varphi_l(s,\tau)=\partial_\tau \Phi_l(s,\tau) \;.
\end{equation}
Analytical progress is again possible for the generic relaxation
scenarios that emerged from the discussion of the bulk in
Sec.~\ref{sec:lin}. We therefore take the initial conditions to be of
the power--law form in Eq.~(\ref{eq:self-affine-inicond}). To simplify
the notation we will from now on drop the subscripts ``$l$'' for
coarse--grained quantities. As before, we consider \emph{type~I} and
\emph{type~II} conditions separately.

\subsection{Exact Similarity Solutions (\emph{Type~II})} 
\label{sec:ess}
\begin{figure}
       \psfrag{x}{$s/\lambda_\beta(\tau)$}
       \psfrag{y}{$\hspace{-.5cm}\sqrt{\alpha_\beta}\chi_\beta$}
       \psfrag{0}{$0$} \psfrag{1}{$1$} \psfrag{3}{$3$}
       \psfrag{b=1.2}{$\beta=1.2$} \psfrag{b=1.6}{$\beta=1.6$}
       \psfrag{b=2}{$\beta=2$} \psfrag{b=2.4}{$\beta=2.4$}
       \psfrag{b=2.8}{$\beta=2.8$} \begin{center}
       \includegraphics[width=\columnwidth]{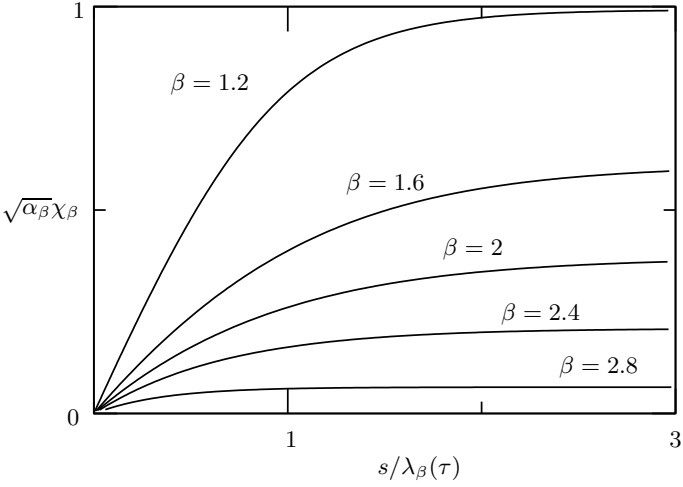}
       \caption{\emph{Type~II:} Stress profile in the boundary layer,
       given by the scaling function
       $\chi_\beta(s/\lambda_\beta(\tau))$,
       Eq.~(\ref{eq:phi-from-chi}). The case $\beta=2$ corresponds to
       thermalized initial conditions.}
       \label{fig:fractal-tension-profiles} \end{center}
\end{figure}

For \emph{type~II} initial conditions, i.e.\
Eq.~(\ref{eq:self-affine-inicond}) with $1<\beta<3$, one can find
exact similarity solutions of
Eq.~(\ref{eq:classical-mech-ode-for-Phi}). To this end, we make the
dynamic scaling ansatz
\begin{equation}
  \label{eq:scaling-ansatz-2}
  \Phi(s,\tau)= \tau^{1/2}\psi_\beta\left(\frac{s}{\lambda_\beta(\tau)}\right) 
\end{equation}
for the integrated force, with the characteristic length
\begin{equation}
  \label{eq:lambda-beta}
  \lambda_\beta(\tau)=\hat{\zeta}^{-1/2}\Lambda^{1-4\delta_\beta}
  \tau^{\delta_\beta} \quad \text{with} \quad \delta_\beta =
  \frac{3-\beta}8 \;.
\end{equation}
In Eq.~(\ref{eq:scaling-ansatz-2}) the bulk dynamics has been
explicitely taken out of the scaling form, and the definition of
$\lambda_\beta$ naturally results from inserting
Eq.~(\ref{eq:scaling-ansatz-2}) into
Eq.~(\ref{eq:classical-mech-ode-for-Phi}) with the aim of eliminating
the parameter dependence.  That the resulting differential equation
for $\psi_\beta(\xi)$ is in particular (essentially) time independent
for $\tau\ll l^4$ is more easily seen after another variable
transformation $q \to \tilde q \tau^{-1/4}$,
\begin{equation}
  \label{eq:ode-chi} \psi_\beta''(\xi)=\mint{d\tilde
  q}{\tau^{1/4}/l}{\infty}\tilde q^{-\beta}\left[e^{2\tilde
  q^2(-\tilde q^2+\psi_\beta(\xi))}-1\right] \;.
\end{equation}
The boundary conditions are $\psi_\beta(0)=\psi_\beta''(\xi \to
\infty)=0$. Having solved Eq.~(\ref{eq:ode-chi}) for $\psi_\beta(\xi)$,
the tension is extracted by differentiation,
\begin{eqnarray}
  \label{eq:phi-from-chi} 
\varphi(s,\tau)&=&\partial_\tau
  \Phi_l(s,\tau) \nonumber \\
  &=&\partial_\tau\left[\tau^{1/2}\psi_\beta\left(\frac{s}{\lambda_\beta(\tau)}\right)\right]
  \nonumber \\
  &\equiv&\sqrt{\frac{\alpha_\beta}{\tau}}\chi_\beta\left(\frac{s}{\lambda_\beta}\right)
  \;.
\end{eqnarray}
To make contact with Eq.~(\ref{eq:ansatz}), the amplitude
$\sqrt{\alpha_\beta}$ calculated in
Sec.~\ref{sec:similarity_solutions} (see
Fig.~\ref{fig:critical-alphas}) was explicitely taken out of the
scaling function, so that the latter is normalized,
$\chi(\xi\to\infty)=1$. The combination
$\sqrt{\alpha_\beta}\chi_\beta(\xi)$ then obeys
\begin{equation}
  \label{eq:identify-alfa}
  \sqrt{\alpha_\beta}\chi_\beta(\xi)=\frac{1}{2}\psi_\beta(\xi)
  -\frac{3-\beta}8\xi\psi_\beta'(\xi)\;.
\end{equation}
In Fig.~\ref{fig:fractal-tension-profiles} the numerical solutions are
shown for different values of $\beta$. We have plotted the combination
$\sqrt{\alpha_\beta}\chi_\beta(s/\lambda_\beta)$ instead of the
normalized scaling function $\chi_\beta$, because the graphs of the
latter cross each other for different $\beta$, rendering the figure
too crowded. The slope $\chi_\beta'(\xi)$ at the origin thus has a
somewhat weaker dependence on $\beta$ as suggested by
Fig.~\ref{fig:fractal-tension-profiles}. It is seen that
$\chi_\beta(s/\lambda_\beta)$ saturates for $s\simeq\lambda_\beta$,
which establishes $\lambda_\beta$ as the characteristic width of the
boundary layer. Fig.~\ref{fig:fractal-tension-profiles} moreover shows
that the tension profiles have a nonzero curvature throughout the
boundary layer. According to Eq.~(\ref{eq:coarse-graining}) the
release of tension and stored length is thus spread over the whole
boundary layer. Observe that $\Lambda^{1-4\delta_\beta} \propto
1/\sqrt{\epsilon^0}$ from Eq.~(\ref{eq:wbLambda}) so that
\begin{equation}\label{eq:lambda_epsilon}
 \lambda_\beta \propto \tau^{\delta_\beta}/\sqrt{\epsilon^0}\;.
\end{equation}
Interestingly, the small parameter $\epsilon^0$ appears in the
denominator so that the limits $\tau\to0$ and $\epsilon^0\to 0$ do not
interchange. This indicates that the boundary layer phenomena are not
accessible by regular perturbation theory.

In the interesting case of thermal initial conditions ($\beta=2$) the
boundary layer $\lambda_{\beta=2}(\tau)$ grows according to
\begin{equation}\label{eq:lambda-2}
  \lambda_2(\tau)=\hat{\zeta}^{-1/2}\Lambda^{-1/2}\tau^{1/8} \;,
\end{equation}
where $2\Lambda/\pi$ is the persistence length.  This particular
relaxation scenario can be imagined to be the consequence of a sudden
temperature jump from finite to zero temperature.  Interestingly, the
boundary layer length $\lambda_2(\tau)$ also governs the thermodynamic
propagation of tension in linear response if a weak longitudinal force
is suddenly applied at one end \cite{everaers-Maggs:99}.

\subsection{Approximate Similarity Solutions (\emph{Type~I})}
\label{sec:ass}

As explained in Sec.~\ref{sec:similarity_solutions}, initial
conditions with roughness $\beta<1$ and large wave number cutoff 
\begin{equation}
  \label{eq:cut-off-condition}
  q_N\gg Q\sim\tau^{-1/4} \;.
\end{equation}
fulfill the criterion Eq.~(\ref{eq:alpha-estimate-integral}) for
\emph{type~I} behavior (Section~\ref{sec:cascade}).  In this case we
can use the right hand side of Eq.~(\ref{eq:I1=I2}) to estimate
$\Delta \epsilon_l(\tau)$ in Eq.~(\ref{eq:coarse-graining}),
\begin{equation}
  \label{eq:adiabatic-equation}
  \hat{\zeta}^{-1}\Phi''\approx\epsilon^0(Q) \Delta Q
  \exp\left(2\alpha\right)-\epsilon^0 \; ,
\end{equation}
where we approximated
\begin{equation}
  \label{eq:stored-length-in-high-modes}
  \mint{dq}{\sqrt{2}Q }{q_N} \epsilon^0(q)\stackrel{q_N\gg Q}{\approx}\mint{dq}{0}{q_N}
  \epsilon^0(q)\equiv\epsilon^0 \;,
\end{equation}
as valid for $\beta<1$. Rather than in the absolute value of $\Phi$ we
are interested in the ratio
\begin{equation}
  \label{eq:psi}
  \hat\psi(s,\tau)\equiv \frac{\Phi(s,\tau)}{\Phi_\infty(\tau)} \;,
\end{equation}
where $\Phi_\infty(\tau)$ is the value of $\Phi$ in the bulk,
\begin{equation}
  \label{eq:Phi-infty}
  \Phi_\infty(\tau)\equiv \lim_{s \to \infty} \Phi(s,\tau) \;.
\end{equation}
By the definitions Eq.~(\ref{eq:psi},~\ref{eq:Phi-infty})
$\hat\psi(s\to\infty,0)=1$ and $\hat\psi(0,\tau)=0$ at the free end to satisfy
the boundary condition. As in Sec.~\ref{sec:cascade}, the vanishing of
the left hand side of Eq.~(\ref{eq:adiabatic-equation}) in the bulk,
\begin{equation}
  \label{eq:bulk-equation-again}
  0\approx\epsilon^0(Q_\infty) \Delta
  Q_\infty\exp\left(2\alpha_\infty\right)- \epsilon^0
\end{equation}
implies that the exponent 
\begin{equation}
  \label{eq:bulk-Phi2}
  2\alpha_\infty\equiv\frac{\Phi_\infty(\tau)^2}{ 2 \tau} \approx \mbox{const.}
\end{equation}
is constant in time up to logarithmic corrections. Now we divide
Eq.~(\ref{eq:adiabatic-equation}) by $\epsilon^0$ and obtain using
Eqs.~(\ref{eq:psi},~\ref{eq:bulk-equation-again},~\ref{eq:bulk-Phi2})
\[
2\frac{\sqrt{\alpha_\infty\tau}}{\hat \zeta
    \epsilon^0}\hat\psi''=1-\frac{\epsilon^0(Q)\Delta
  Q\exp[2\alpha]}{\epsilon^0(Q_\infty) \Delta
  Q_\infty\exp[2\alpha_\infty]} \;. 
\] 
Inserting the initial conditions Eq.~(\ref{eq:self-affine-inicond})
and using the definitions of $\alpha$, $Q$ and $\Delta Q$ from
Sec.~\ref{sec:ampelfaktor}, we arrive at
\begin{equation}
  2\frac{\sqrt{\alpha_\infty\tau}}{\hat \zeta
    \epsilon^0}\hat\psi''=1-\hat\psi^{-\frac{1+\beta}{2}}\exp\left[2
    \alpha_\infty (\hat\psi^2-1)\right] \label{eq:eq-giving-b-of-t} \;.
\end{equation}
For given $\alpha$ Eq.~(\ref{eq:eq-giving-b-of-t}) is solved by the
scaling ansatz
\begin{equation}
  \label{eq:scalingform-peaked-solutions}
  \hat\psi(s,\tau)= \hat\psi\left(\frac{s}{\lambda(\tau)}\right) \;,
\end{equation}
where the width $\lambda(\tau)$ of the boundary layer is now given by
\begin{equation}
  \label{eq:b-of-t}
  \lambda(\tau)\equiv 2 (\hat{\zeta}\epsilon^0)^{-1/2} (\tau
  \alpha_\infty )^{1/4} \;,
\end{equation}
and thus --- in contrast to what we found under \emph{type~II}
conditions --- is directly proportional to $Q_\infty^{-1}(\tau)$.  As
the boundary layer width $\lambda_\beta$ under \emph{type~II}
conditions, it is inversely proportional to $\sqrt{\epsilon^0}$, which
entails the same conclusions as drawn after
Eq.~(\ref{eq:lambda_epsilon}).

\begin{figure}
       \psfrag{x}{$s/b$} \psfrag{y}{$\chi$} \psfrag{0}{$0$}
       \psfrag{1}{$1$} \psfrag{2}{$2$} \psfrag{4}{$4$}
       \psfrag{0,4}{$0.4$} \psfrag{0,8}{$0.8$} 
       \psfrag{increasing a}{increasing $\alpha_\infty$}  
       \psfrag{b=-5}{$\beta=-5$}
       \psfrag{b=-11}{$\beta=-1$} \psfrag{b=.5}{$\beta=.5$}
       \begin{center}
       \includegraphics[width=\columnwidth]{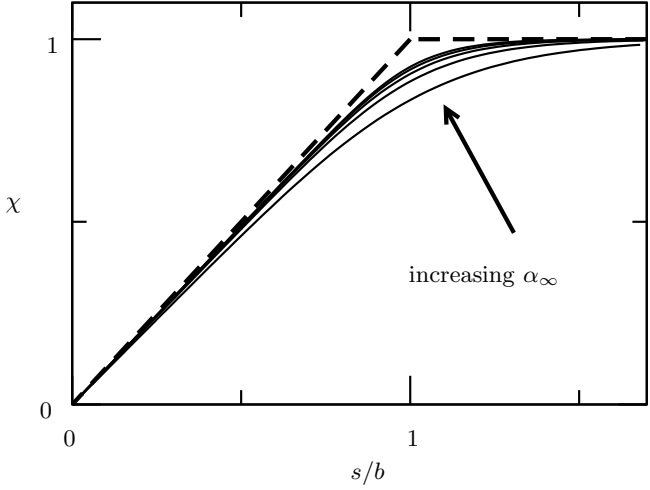}
\caption{\emph{Type~I}: Stress profile in the boundary layer, given by
the scaling function $\chi(s/\lambda_\beta(\tau))$,
Eq.~(\ref{eq:tensionfromchipeaked}). The displayed curves correspond
to $\beta=-1$ and $\alpha_\infty=5,10,15,20,25$. For increasing
$\alpha_\infty\gg 1$, $\chi(\xi)$ approaches its limiting form
$\chi(\xi<1)=\xi$ and $\chi(\xi>1)=1$. The asymptotic behavior is
independent of $\beta<1$.}
\label{fig:peaked-tension-profiles} \end{center}
\end{figure}

Inserting the scaling form Eq.~(\ref{eq:scalingform-peaked-solutions})
into Eq.~(\ref{eq:eq-giving-b-of-t}) yields
\begin{equation}
  \label{eq:chi-equation} \frac{1}{2}\hat\psi''(\xi)= 1 -
  \hat\psi^{-\frac{1+\beta}{2}}\exp\left[\alpha_\infty
  (\hat\psi^2-1)\right]\;.
\end{equation}
After solving Eq.~(\ref{eq:chi-equation}) for $\hat\psi(\xi)$ the tension is found as before,
\begin{eqnarray}
  \label{eq:tensionfromchipeaked}
  \varphi(s,\tau)&=&\partial_\tau\Phi(s,\tau) \nonumber \\ &=&
  \sqrt{\frac{\alpha_\infty}{\tau}}
  \left(\hat\psi\left(\frac{s}{\lambda}\right)-\frac{s}{2
  \lambda}\hat\psi'\left(\frac{s}{\lambda}\right)\right) \nonumber \\
  &=& \sqrt{\frac{\alpha_\infty}{\tau}}
  \chi\left(\frac{s}{\lambda(\tau)}\right) \;,
\end{eqnarray}
where the normalized scaling function $\chi(\xi)$ is given by
\begin{equation}
  \label{eq:chi}
  \chi(\xi)=\hat\psi(\xi)-\frac{1}{2}\xi \hat\psi'(\xi)\;. 
\end{equation}
In Fig.~\ref{fig:peaked-tension-profiles} the scaling function
$\chi(\xi)$ is shown for different $\alpha_\infty>1$. With increasing
$\alpha_\infty$ the curves converge from below to a piece--wise linear
form that consists of a linear boundary layer $\chi(\xi)=\xi$ for
$\xi<1$ and a bulk area $\chi(\xi)=1$ for $\xi>1$. This limiting
behavior is independent of the roughness exponent $\beta<1$. According
to Eq.~(\ref{eq:coarse-graining}) it corresponds to a completely
straightened boundary layer with the linearly growing tension being
fully due to the accumulating force from the viscous friction against
the solvent, and a buckled bulk regime with a spatially constant
pressure conserving its initially stored length. From
Eq.~(\ref{eq:coarse-graining}) it is moreover seen that the limit
$\alpha_\infty\to\infty$ physically corresponds to a situation where
the region of stored length release shrinks to a single point at
$s=\lambda(\tau)$ that separates the buckled bulk from the relaxed
boundary layer.  This is in accord with our conclusion at the end of
Sec.~\ref{sec:lin} that for \emph{type~II} initial conditions tension
decay and conformational relaxation can be identified (see
Fig.~\ref{fig:terminal-relaxation}).

{F}rom what was said there, we therefore could have guessed the results
of the preceding paragraph from an intuitive scaling argument that
reverses the above line of arguments. Starting from the very
assumption that the rod consists of totally straightened tails of
length $\lambda(\tau)$ that dynamically constrain the bulk, one
concludes that the bulk pressure $\varphi_\infty(\tau)$ drives the
tails outwards at the velocity $v_\parallel$ needed to balance this
pressure by the Stokes friction onto the tails, i.e.
\begin{equation}\label{eq:scaling-law-boundary-layer}
  \varphi_\infty(\tau)=\hat \zeta v_\parallel \lambda(\tau)\;.
\end{equation}
Given the straight contour of the moving part $\lambda(\tau)$ this
then entails the linear tension profile within the boundary layer.  On
the other hand, the velocity of the boundary layer must be equal to
the stored length release per unit of time
$v_\parallel=\epsilon^0\partial_\tau \lambda $, which takes place in
the small crossover region between bulk and boundary layer. Using the
power law Eq.~(\ref{eq:ft}) for the pressure within the (constrained)
bulk with a prefactor $\sqrt{\alpha_\infty}$ we obtain the closed
differential equation
\begin{equation}\label{eq:scaling-arg-for-b}
 \epsilon^0 \hat \zeta \lambda\partial_\tau \lambda=
    \sqrt{\alpha_\infty/\tau} \;,
\end{equation}
which is solved by expression Eq.~(\ref{eq:b-of-t}) for the length
$\lambda(\tau)$.

\subsection{Consistency} 
Since both lengths $Q^{-1}_\infty(\tau)$ and $\lambda(\tau)$ grow in
time --- under \emph{type~II} conditions even with different exponents
--- one might worry about the time domain of validity of the length
scale separation Eq.~(\ref{eq:adiabatic_length}) underlying the above
derivation. Consistency of the adiabatic approach requires that the
wavelength $Q_\infty^{-1}= (\tau/\alpha_\infty)^{1/4}$ that dominates
the sum over all modes is much smaller than the length over which the
tension varies, i.e. the width of the boundary layer $\lambda(\tau)$.
For \emph{type~II} initial conditions, we thus need
\begin{equation}
  \label{eq:selfconsistency2} Q_\infty(\tau)
  \lambda_\beta(\tau)=\alpha_\beta^{1/4}\hat \zeta^{-1/2}
  \left(\Lambda/\tau^{1/4}\right)^{\beta-1} \gg 1\;
\end{equation}
For $\tau\rightarrow 0$ the inequality Eq.~(\ref{eq:selfconsistency2})
is certainly true, because $\beta-1>0$. The product
$Q_\infty\lambda_\beta$ becomes comparable to one for
$\lambda_\beta(\tau)\approx\Lambda$. However, this point of
inconsistency cannot be reached, since we had to assume in the
discussion after Eq.~(\ref{eq:self-affine-inicond}) that $L\ll
\Lambda$ in order to ensure the weakly bending limit.  Likewise, for
the \emph{type~II} scenario we need
\begin{equation}
  \label{eq:selfconsistency1}
  Q_\infty(\tau) \lambda(\tau)=2\left[\alpha_\infty/\left(\hat \zeta
      \epsilon^0\right)\right]^{1/2} \gg 1 \;.
\end{equation}
Again, this generally holds in the weakly bending limit
Eq.~(\ref{eq:epsilon}).

The adiabatic approximation thus proves to be able to describe the
arclength dependent tension relaxation in the weakly bending limit. On
the other hand, the consistency conditions
Eqs.~(\ref{eq:selfconsistency2},~\ref{eq:selfconsistency1}) can be
taken as another indication that the weakly bending limit is in fact a
necessary ingredient for the universality of the relaxation process
and in particular for the characteristic power--law relaxation
Eq.~(\ref{eq:ft}).

\subsection{Terminal Relaxation} 
\label{sec:term-rel}
Up to now, we have considered the growth of the boundary layer in a
rod that has a (formally) semi--infinite arc length parameter space,
$s=0\dots\infty$, which is an idealization. However, the foregoing
discussion obviously applies equally to a free rod of \emph{finite}
length $L$ for sufficiently short times: As long as the size of the
boundary layer is much smaller than the total length $L$ the presence
of a second free end is irrelevant to the boundary layer at the first
end.  The time where the boundary layers span the whole rod marks the
crossover to a new behavior. For definiteness, we define the crossover
time $\tau_f$ by
\begin{equation}
  \label{eq:tension-rel-time}
  \lambda(\tau_f)\equiv L \;.
\end{equation}
Further contour relaxation proceeds essentially free of lateral
stress, because the tension is equilibrated everywhere with the free
ends for $\tau\gg\tau_f$. The time $\tau_f$ can thus be identified as
the characteristic decay time for the tension.  From
Eqs.~(\ref{eq:greens_fct},~\ref{eq:amplific-fac}) it is seen that
after time $\tau_f$ all modes decay exponentially throughout the whole
rod,
\begin{equation}
  \label{eq:zero-tension-relaxation}
  \epsilon(q,\tau > \tau_f)\approx 
 \epsilon(q,\tau_f)e^{-q^4 (\tau-\tau_f)} \;.
\end{equation}
Stored length is no longer conserved and a mode with wave number $q$
has thus decayed after time $\tau\approx \tau_f+q^{-4}$.

We recall from our discussion at the end of Sec.~\ref{sec:lin} that
Eq.~(\ref{eq:zero-tension-relaxation}) corresponds to very different
behavior of the overall conformational relaxation for \emph{type~I}
and \emph{type~II} initial conditions, respectively. Consider again
Fig.~\ref{fig:terminal-relaxation}. In the \emph{type~I} scenario, at
$\tau_f$ the stored length is concentrated in modes with wavelengths
$q\approx Q^{-1}(\tau_f)$, which (visibly) dominate the contour
undulations. Hence, we can conclude that it takes a time of the order
of $\tau_f$ to release the bulk of the initially stored length
$\epsilon^0(0)$ after the dynamic confinement ceases. In other words,
the rod straightens within a time of the order of $\tau_f$. This
conforms with the earlier conclusions that tension relaxation and
stored--length release occur in parallel, so that the conformation in
the boundary layer is the straight ground state.

On the contrary for \emph{type~II} initial conditions the stored
length distribution in mode space hardly differs from the initial
condition Eq.~(\ref{eq:self-affine-inicond}), i.e.\ it is still
strongly peaked at low $q$ and the total stored length has not changed
appreciably. It takes a time $\tau=L^4$ until the contour undulations
that carry most of the stored length have relaxed.  From the
definition, Eq.~(\ref{eq:tension-rel-time}), of $\tau_f$ and the
boundary layer growth law, Eq.~(\ref{eq:lambda-beta}), we infer
\begin{equation}
  \label{eq:tf-vs-tl}
  L^4/\tau_f= (\epsilon^0)^{-1/(2\delta_\beta)}\gg1 \;.
\end{equation}
The conformational relaxation takes much longer than $\tau_f$,
particularly as $\delta_{\beta\to 3}\to0$. This heralds the (trivial)
limit of instant tension equilibration for $\beta\geq 3$. As already
observed for a confined rod at the end of Sec.~\ref{sec:lin} as well
as for the boundary layers discussed in Sec.~\ref{sec:ess}, the
conformational relaxation for $\beta>1$ lags behind the stress
relaxation.

We finally comment on the relation to the regular perturbation
approach of Sec.~\ref{sec:lin}. For free ends, it would to lowest
order predict
\begin{equation}
  \label{eq:reg-pert-1}
  \varphi(s,\tau)=\mbox{const.}=0 \;.
\end{equation}
At first sight this contradicts our intuitive understanding that the
bulk of a relaxing rod should be under pressure at least for short
times. However, fixing the length and time scales $L$ and $\tau$ of
the problem while $\epsilon^0\to 0$, the prediction of zero tension is
indeed recovered from the adiabatic approach via the vanishing of the
relaxation time $\tau_f$ in Eq.~(\ref{eq:tension-rel-time}). This is
apparent from the above Eq.~(\ref{eq:tf-vs-tl}) for \emph{type~I}
initial conditions and from Eq.~(\ref{eq:b-of-t}) for \emph{type~II}
initial conditions.  Thus for any fixed given total length $L$ and
time $\tau$ there exists an $\epsilon^0_c$ such, that the prediction
of the multiple scale perturbation theory reduces to that of the
regular perturbation scheme for $\epsilon^0\ll\epsilon^0_c$. However,
the interesting short--time regime $\tau\ll\tau_f(\epsilon^0,L)$ is
not accessible by regular perturbation theory. Upon fixing
$\epsilon^0$ and $L$ and considering small $\tau \to 0$ (i.e.\ the
situation just after removing the confining walls that served to keep
the tension spatially constant), the decay of the bulk tension
obviously has to occur in an arbitrarily narrow boundary region. In
other words, the putative $O(\epsilon)-$term $f'$ in
Eq.~(\ref{eq:parallel-eq-motion}) has to diverge on physical grounds,
thus signaling the breakdown of regular perturbation theory for open
boundary conditions in this limit.

\section{Comparison with Numerical Simulations}
\label{sec:simulation}

In this section we want to point out how our results for the stress
relaxation manifest themselves in various observables that have been
monitored in numerical simulations.  Golubovic et
al.~\cite{golubovic-moldovan-peredera:98} investigated the effect of a
sudden temperature jump on an initially straight rod of length $l$
confined between two walls (hinged ends). The frustration due to
thermal expansion is modeled by a relative initial compression
$\epsilon^0\ll1$ of the backbone of the rod.  The consequent initial
pressure $\varphi_i$ drives the evolution of buckles with wave number
$Q_i=\sqrt{\varphi_i/2}$. After a transition period characterized by
backbone expansion most of the length $\epsilon^0 l$ is stored in
bending modes with wave number close to $Q_i$ rather than in backbone
vibrational modes. The backbone length appears to be almost constant
from thereon. The rod relaxes in this second stage as if it was
incompressible with a pronounced peak in the initial mode spectrum
(prepared by the thermal expansion of the rod).  The scenario thus
agrees with the assumptions of Sec.~\ref{sec:cascade}. Our analysis
there explains why and how the peak grows and sharpens in time.
Asymptotically, we predict the dominant wave number to evolve
according to $Q\propto \tau^{-1/4}$, as observed in
Ref.~\cite{golubovic-moldovan-peredera:98} by analyzing the
tangent--tangent correlation function. The fundamental scaling law
Eq.~(\ref{eq:ft}) for the tension derived in Sec.~\ref{sec:cascade} is
the basis for the power--law time evolution of a number of other
observables.  For example, the mean--square transverse displacement
\begin{equation}\label{eq:w2}
w^2 \equiv
l^{-1}\mint{ds}{0}{l}r_\perp^2(s,\tau)
\end{equation}
was observed to obey $w^2=2 \epsilon^0 \tau^{1/2}\propto \tau^{1/2}$
\cite{golubovic-moldovan-peredera:98} and interpreted as an immediate
consequence of the existence of a dominant wave length, which we
established above for \emph{type~I} initial conditions.  From the
dominance of $Q$ together with the conservation of stored length, one
has
\begin{eqnarray}
  \label{eq:golubo1}
  w^2&=&2\sum_n \epsilon(q_n,\tau) q_n^{-2} \nonumber\\
  &\approx& 2\epsilon(Q,\tau)Q^{-2}  \nonumber \\
  &=& 2 \epsilon^0 Q^{-2} 
 \sim\tau^{1/2}  \;.
\end{eqnarray}
Analogous arguments can be used for other observed quantities, such as
the stored elastic energy or the dissipation rate etc.  In particular,
as we have shown in Sec.~\ref{sec:ampelfaktor},~\ref{sec:cascade}, the
cascading of stored length in mode space maintains and enhances the
maximum in the mode spectrum asymptotically, even if the initial mode
spectrum is sufficiently slowly decaying.  The validity of
Eq.~(\ref{eq:golubo1}) and related scaling behavior in other
observables thus also extend to this situation.

The simulations by Spakowitz and Wang \cite{spakowitz:2001} considered
the same setup as in Ref.~\cite{golubovic-moldovan-peredera:98} but
with free boundary conditions. Besides a higher order effect guiding
the evolution of helical modes, the same power laws are found for the
dominant wave number and the evolution of transverse displacements,
respectively. This becomes better comprehensible from our boundary
layer calculations, which show that most of the rod should indeed
behave as if it were longitudinally confined as long as the boundary
layer does not span the whole filament, e.g.\ for $\tau\ll\tau_f$.
Additionally, as a measure for the longitudinal expansion Spakowitz
and Wang \cite{spakowitz:2001} proposed a longitudinal radius of
gyration $R^G_\parallel$ as the largest eigenvalue of a gyration
tensor. In our terms, this quantity can be identified with
\begin{equation}
  \label{eq:gyration-radius}
  R^2_{G\parallel}(\tau)\equiv \frac{1}{L}\mint{ds}{0}{L}
  \left[z_{CM}-s+r_\parallel(s,\tau)\right]^2
\end{equation}
for a rod with a time independent longitudinal center of mass
coordinate $z_{CM}=s_{CM}-r_\parallel(s_{CM})$ lying approximately at
the center of the rod,
\begin{equation}
  \label{eq:CM}
  z_{CM}+r_\parallel(0)\approx L/2 \;.
\end{equation}
Using the arguments developed above, the time derivative
$\partial_\tau R^2_{G\parallel}(\tau)$ is in the limit $\tau\to 0$
given by
\begin{subequations}\label{eq:dtrg2}
  \begin{align}
    R_{G\parallel} \partial_\tau R_{G\parallel} &= \frac{1}{L}
    \mint{ds}{0}{L} \left[z_{CM}-s+r_\parallel
    \right]_{\tau=0}\partial_\tau r_\parallel \nonumber \\
    &\approx  \mint{ds}{0}{L/2}\partial_\tau r_\parallel
   \label{eq:rg-est1} \\
   &= \varphi_\infty \label{eq:rg-est2} \propto \tau^{-1/2} \;.
  \end{align}
\end{subequations}
The first approximation Eq.~(\ref{eq:rg-est1}) follows from
Eq.~(\ref{eq:CM}) and from the fact that for short times
$\partial_\tau r_\parallel$ is finite (to leading order) only close to
the ends, $s=0$ and $s=L$. The subsequent Eq.~(\ref{eq:rg-est2}) holds
because of Eq.~(\ref{eq:leading-order-tension-eq}). Integrating
Eq.~(\ref{eq:dtrg2}) in time and observing $R^G_\parallel(0)\approx
L/(2\sqrt 3)$ one gets the algebraic growth law
\begin{equation}\label{eq:rgt}
\delta R^G_\parallel(\tau) \equiv 
R^G_\parallel(\tau)-R^G_\parallel(0)\propto\tau^{1/2} \;,
\end{equation}
which is indeed empirically found to hold with high accuracy over a
broad time window \cite{spakowitz:2001}. Note, however, that according
to Eq.~(\ref{eq:dtrg2}) the (initial) variation of the radius of
gyration measures the time integral of the bulk tension rather than
the growth of the boundary layer. It thus provides a practical direct
measure of $\Phi(\tau)$, but is not suitable to monitor the
conformational relaxation. 

Access to the latter can be gained by probing the end--to--end
distance
\begin{equation}\label{eq:end_to_end}
R_\parallel=L-r_\parallel(L)+r_\parallel(0)
\end{equation}
instead. Its temporal change $\delta R_\parallel(\tau)\equiv
R_\parallel(\tau)-R_\parallel(0)$ is obviously directly due to
stored--length release. Under \emph{type~I} conditions, where
stored--length release and tension decay go hand in hand and the
boundary layer is essentially straight, the released length is nothing
but the total stored length that was initially contained in the
boundary layer, i.e.
\begin{equation}\label{eq:rpar}
  \delta R_\parallel(\tau)\approx \epsilon^0 
  \lambda(\tau)\sim\sqrt{\epsilon^0}\tau^{1/4}\;.
\end{equation}

Rods with \emph{type~II} initial conditions behave differently. Again,
stored--length release does not occur in the bulk.  However, the
stored length in the boundary layer is released much slower than the
boundary layer grows, as discussed in Sec.~\ref{sec:term-rel}. Not all
of the initially stored length but only some fraction
$\Delta\epsilon^*(q,\tau)$ has been released after time $\tau$.  The
latter can be estimated from Eq.~(\ref{eq:zero-tension-relaxation}),
since the relaxation within the boundary layer is essentially tension
free:
\begin{eqnarray}
  \label{eq:zero-tension-relaxation-2}
  \Delta\epsilon^*(q,\tau)&=&\mint{dq}{L^{-1}}{\infty}
  \epsilon^0(q)\left(e^{-q^4\tau}-1\right) \nonumber \\
  &\stackrel{\tau^{1/4}\ll
  L}{\sim}&\Lambda^{1-\beta}\tau^{\frac{\beta-1}4} \;.
\end{eqnarray}
Asymptotically we can thus write
\begin{equation}\label{eq:rpar2}
\delta R_\parallel(\tau)\approx \Delta\epsilon_\beta^*(\tau)
\lambda_\beta(\tau) \propto
\sqrt{\epsilon^0}\tau^{\frac{\beta-1}4+\delta_\beta} \;,
\end{equation}
for short times $\tau\ll\tau_f$. In the last step we used
$\Lambda^{1-\beta}\propto\epsilon^0$ from Eq.~(\ref{eq:wbLambda}).
For  $\tau\approx \tau_f$ the growth of the boundary layer saturates at
$\lambda \simeq L$, so that for long times $\tau\gg\tau_f$
\begin{equation}\label{eq:rpar3}
\delta R_\parallel(\tau)\approx \Delta\epsilon_\beta^*(\tau) L \propto
\epsilon^0 \tau^{\frac{\beta-1}4}\;.
\end{equation}
In summary, for \emph{type~II} initial conditions
\begin{equation}\label{eq:rparsum}
    \delta R_\parallel(\tau) \propto 
 \begin{cases}
  \sqrt{\epsilon^0}\tau^{\rho_\beta^<} , \, \quad \rho_\beta^< =
  \frac{\beta+1}8  & (\tau\ll\tau_f) \\
  \epsilon^0 \tau^{\rho_\beta^>} , \qquad \rho_\beta^>=  \frac{\beta-1}4
 &  (\tau\gg\tau_f) \;.
 \end{cases}
\end{equation}
In particular, we note that an initially thermalized rod first
expands according to $\delta R_\parallel\propto \tau^{3/8}$, and
eventually as $\delta R_\parallel\propto \tau^{1/4}$. The exponents
$\rho_\beta^<$ and $\rho_\beta^>$, which obey $\rho_\beta^<=
\rho_\beta^>+\delta$, are displayed in Fig.~\ref{fig:delta-beta}
together with $\delta_\beta$ for comparison.

The initial growth laws Eqs.~(\ref{eq:rpar},~\ref{eq:rpar2}) including
prefactors can also be derived more rigorously from the scaling forms
for the integrated tension derived in
Secs.~\ref{sec:ess},~\ref{sec:ass}.  For \emph{type~I} initial
conditions $\Delta \epsilon(s,\tau)/\epsilon^0$ is given by the
right--hand--side of Eq.~(\ref{eq:eq-giving-b-of-t}), hence
\begin{equation}\label{eq:del-r-exact2}
 \begin{split}
   \delta R_\parallel(\tau)&=2\mint{ds}{0}{\infty}\Delta\epsilon(s,\tau) \\
   &=2\mint{ds}{0}{\infty}\epsilon^0
   \left\{1-\hat\psi^{-\frac{1+\beta}{2}}\exp\left[2 \alpha_\infty
       (\hat\psi^2(s,\tau)-1)\right]\right\}
   \\
   &=\lambda(\tau) \epsilon^0 
   \mint{d\xi}{0}{\infty}\left\{1-\hat\psi^{-\frac{1+\beta}{2}}\exp\left[2
       \alpha_\infty (\hat\psi^2(\xi)-1)\right]\right\} \\
   &\propto \sqrt{\epsilon^0} \tau^{1/4}\;.
  \end{split}
\end{equation}
The exact prefactor can be obtained as a function of the appropriate
scaling function by evaluating the integral numerically. For
\emph{type~II} initial conditions one finds
\begin{equation}\label{eq:del-r-exact}
 \begin{split}
   \delta R_\parallel(\tau)&=2\mint{ds}{0}{\infty}\Delta\epsilon(s,\tau) \\
   &=2\mint{ds}{0}{\infty}\Lambda^{1-\beta}\mint{dq}{0}{\infty}
   q^{-\beta}\left[e^{2q^2\left[\Phi(s,\tau)-q^2\tau\right]}-1\right]
   \\
   &=\Lambda^{1-\beta}\tau^{\frac{\beta-1}4}\lambda_\beta(\tau) \times
   \\
   &\quad \mint{d\xi}{0}{\infty} 2 \mint{dq}{0}{\infty}
   q^{-\beta}\left[e^{2q^2\left[\psi(\xi)-q^2\right]}-1\right]\\
   &\propto \sqrt{\epsilon^0} \tau^{\rho_\beta}\;,
  \end{split}
\end{equation}
which also gives an explicit expression for the prefactor in terms
of the scaling function $\psi(\xi)$.
 
We finally comment on a possible problem that could arise because of
the 'microscopic' nature of $\delta R_\parallel$. Note that in
contrast to $\delta R^G_\parallel$ it is also sensitive to microscopic
details of the relaxation and the initial conditions, since it
contains contributions from Fourier modes beyond those corresponding
to the coarse--graining length $l$. In particular, the longitudinal
projection of transverse fluctuations near the ends could possibly mix
into the genuinely longitudinal dynamics, thereby affecting the
observed time dependence.  This effect plays indeed an important role
for the linear response of stiff polymers longitudinally pulled at
their ends \cite{everaers-Maggs:99}. The situation is somewhat more
fortunate in the present case, since one can show the 'microscopic'
contributions to obey the same power--law dynamics but with a
prefactor of lower order in $\epsilon$.

Altogether, it appears that evidence for the bulk relaxation of
\emph{type~I}, i.e.\ in the regime of mode--space localization
corresponding to a roughness exponent $\beta<1$, can be found in
existing simulations. The more complicated \emph{type~II} intermediate
asymptotic regime and our predictions for the boundary layer dynamics
represent interesting new features, which could be verified in
simulations by probing the growth of the end--to--end distance.
Finally, semiflexible polymers may lend themselves to an experimental
investigation of our predictions for the particular case $\beta=2$.

\section{Conclusions and Outlook}
\begin{figure}
  \psfrag{x}{$\beta$} \psfrag{d}{$\delta_\beta$}
  \psfrag{r<}{$\rho^<_\beta$} \psfrag{r>}{$\rho^>_\beta$}
  \psfrag{1/8}{$1/8$} \psfrag{1}{$1$} \psfrag{1/4}{$1/4$}
  \psfrag{3/8}{$3/8$} \psfrag{1/2}{$1/2$} \psfrag{3}{$3$}
  \psfrag{0}{$0$}
  \begin{center}
    \includegraphics[width=\columnwidth]{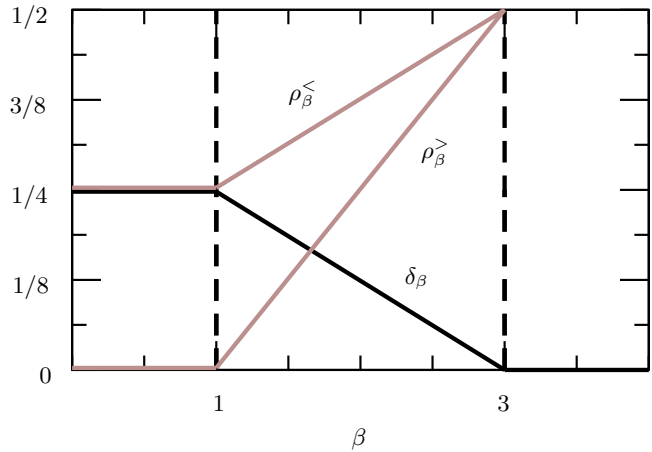}

    \caption{The exponents $\rho^<_\beta$ and $\rho^>_\beta$ (grey) determine 
      the growth $\delta R_\parallel\propto \tau^{\rho_\beta}$ for the
      end--to--end distance on short ($\tau\ll\tau_f$) and long times
      ($\tau\gg\tau_f$), respectively. The exponent
      $\delta_\beta=\rho^<_\beta-\rho^>_\beta$ (black) characterizes
      the growth $\lambda_\beta\sim\tau^{\delta_\beta}$ of the width
      of the boundary layer.}  \label{fig:delta-beta}
      \end{center}
\end{figure}

We have developed and applied in Sec.~\ref{sec:free-relaxation} and
App.~\ref{sec:homogenization} an adiabatic method to calculate the
overdamped heterogeneous stress relaxation in a multiply but weakly
buckled rod. The possible generic relaxation scenarios could
conveniently be characterized in terms of a {\em roughness} exponent
$\beta$ parameterizing the initial excitation. The coarse--grained
pressure $\varphi_l(s,\tau)$ along the rod backbone could be cast into
the universal scaling form
\begin{equation}
  \label{eq:main-result}
 \varphi_l(s,\tau)= \sqrt{\frac{\alpha_\beta}{\tau}}\;
 \chi_\beta\left(\frac{s}{\lambda_\beta(\tau)}\right)\;,
\end{equation}
with a boundary layer width
\begin{equation}\label{eq:main-result2}
\lambda_\beta(\tau)\sim \tau^{\delta_\beta} \;, 
\end{equation}
and a normalized monotone scaling function $\chi_\beta(\xi)$
\begin{equation}
  \label{eq:bc-scaling-form} \chi_\beta(\text{end})=0 \;,\;
  \chi_\beta(\text{bulk})=1 \nonumber \;.
\end{equation}
The latter was calculated numerically and displayed in
Figs.~\ref{fig:fractal-tension-profiles},~\ref{fig:peaked-tension-profiles}
for the two fundamentally different cases $\beta<1$ (\emph{type~I})
and $1<\beta<3$ (\emph{type~II}), respectively.  The amplitude
$\sqrt{\alpha_\beta}$ of the power--law decay of the tension in the
bulk depends on the initial conditions as summarized by
Fig.~\ref{fig:critical-alphas}. The exponent $\delta_\beta$ depicted
in Fig.~\ref{fig:delta-beta} characterizes the growth of the width of
the boundary layer over which the tension continuously decays from its
bulk value $\sqrt{\alpha_\beta/\tau}$ to zero. The particular case of
thermal initial conditions corresponds to
$\sqrt{\alpha_{\beta=2}}\approx 0.386$ and $\lambda_{\beta=2}=1/8$.
For \emph{type~I} initial conditions the dynamics is governed by a
unique characteristic dynamic length scale $Q^{-1}$. Tension
propagation coincides with contour relaxation. The contour has relaxed
by the time $\tau_f$ when the tension has equilibrated throughout the
rod. On the contrary, for \emph{type~II} initial conditions, the
boundary layer width constitutes an additional dynamic length scale
that behaves different from $Q^{-1}$. Tension relaxation precedes
contour relaxation and most of the contour relaxation occurs under
vanishing tension. 

From these central results we derived corresponding power--laws for a
number of observables that seem well suited to test our predictions in
simulations. In particular, we showed that the longitudinal radius of
gyration $R_\|^G$ is suitable to directly probe the (universal)
tension relaxation in the bulk, i.e., the prefactor in
Eq.~(\ref{eq:main-result}). The more complex boundary layer growth,
Eq.~(\ref{eq:main-result2}), which sensitively depends on the type of
initial conditions, was shown to be reflected in the conformational
dynamics. It thus can be accessed by a measurement of the longitudinal
end--to--end distance $R_\|$, which was predicted to exhibit the
intriguing dynamical crossover behavior in Eq.~(\ref{eq:rparsum}).

Following Spakowitz and Wang \cite{spakowitz:2001}, an interesting
route for future theoretical investigations could be to allow for
higher order contributions to the harmonic wormlike--chain Hamiltonian
in Eq.~(\ref{eq:wlc}) to analyze the intriguing non--linear phenomenon
of helix formation and coarsening.

With minor modifications the adiabatic method developed here can be
used to determine stress profiles for non--deterministic, thermal
dynamics (i.e.\ for semiflexible polymers in various situations of
external driving), and will thus be helpful in establishing a unified
description of tension propagation in stiff polymers. In fact, even
the athermal case considered here, can for the special choice
$\beta=2$ be interpreted as special non--equilibrium thermodynamics
problem: the free contour relaxation after a sudden temperature jump
in the limit of vanishing final temperature. The above derived scaling
behavior (but not the amplitudes) can be shown to generalize to the
case that the final temperature is finite
\cite{hallatschek-frey-kroy:04}.

\begin{acknowledgments}
  We gratefully acknowledge helpful discussions with Jan Wilhelm
  during the early stages of this work.
\end{acknowledgments}

\appendix

\section{Method of Homogenization}
\label{sec:homogenization}
Given the separation of the length scales $Q^{-1}$ and $\lambda$
observed in Sec.~\ref{sec:adi}, it is natural to apply the method of
multiple scales \cite{holmes:95} to find an approximate closed
equation for the slow variation of the tension $\varphi(s,\tau)$ over
the length scale $\lambda$ that is independent of the detailed
``microscopic'' fluctuations on the scale $Q^{-1}$. To this end, we
introduce a rapidly and a slowly varying arc length coordinate,
$x\equiv s$ and $y\equiv s \epsilon^{\alpha}$, respectively, where the
exponent $\alpha>0$ will be fixed later.  (The small parameter
$\epsilon\ll1$ will be identified with the fraction of initially
stored length denoted by $\epsilon^0$ above.) Any function $g(s)$
depending on the arc length $s$ is now considered to depend on both
variables $g(s)\to g(x,y)$, where $x$ and $y$ are treated as
independent. The original arclength derivative then becomes
\begin{equation}
  \label{eq:spatio-derivative}
  \partial_s|_\tau \equiv \partial_x|_{\tau,y} + \epsilon^\alpha
  \partial_y|_{\tau,x} \;.
\end{equation}
The dynamic variables $r_\perp$ and $f=\kappa\varphi$ in the equations
of motion Eqs.~(\ref{eq:EOM}) are assumed to have a uniform power
expansion (the expansion coefficients in each order have to be
bounded~\cite{holmes:95}) in terms of the small parameter $\epsilon$,
\begin{eqnarray}
  \label{eq:power-series}
   \vec r_\perp&=&\epsilon^{1/2} \vec h_0+ o(\epsilon^{1/2}) \;, \nonumber \\
   \varphi &=& \varphi_0+\epsilon \varphi_1+o(\epsilon) \;.
\end{eqnarray}
Eliminating the (dependent) coordinate $r_\parallel$ via the local
constraint Eq.~(\ref{eq:local-const}) to the required order and
inserting the power expansions Eq.~(\ref{eq:power-series}) in the
equations of motion yields
\begin{subequations}\label{eq:EOM-ordered}
  \begin{align}
    0=&\,\epsilon^{1/2}\left[\partial_\tau \vec h_0+\partial_x^4\vec
      h_0+\partial_x(\varphi_0 \partial_x \vec
      h_0)\right]+o(\epsilon^{1/2}) \label{eq:EOM-perp-ordered} \\
    0=&\, \partial_x^2 \varphi_0 +\epsilon^\alpha 2 \partial_x\partial_y
    \varphi_0 +\epsilon^{2\alpha}\partial_y^2 \varphi_0 \nonumber\\ &
    +\epsilon\left[\partial_x^2\varphi_1-X_0(x,y)\right]
    \nonumber \\ &+ o(\epsilon;~\epsilon^{2\alpha}) \;.
    \label{eq:EOM-parallel-ordered}
  \end{align}
\end{subequations}
By
\begin{multline} 
\label{eq:nl}
X_0(x,y)=\frac{\hat \zeta}{2}\partial_\tau(\partial_x \vec
h_0)^2+\frac{1}{2}\partial_x^2\left[ \varphi_0\left(\partial_x\vec
    h_0\right)^2\right] \\ +\frac{1}{2}\partial_x^4\left(\partial_x
  \vec h_0\right)^2 \nonumber -(1-\hat
\zeta)\partial_x\left[(\partial_x \vec h_0) (\partial_\tau\vec
  h_0)\right]
\end{multline}
we have summarized terms non--linear in $\vec h_0$. The $O(1)$ part of
Eq.~(\ref{eq:EOM-parallel-ordered}) together with the requirement of
$\varphi_0$ being bounded for large $x$ implies that
\begin{equation}
  \label{eq:x-indep-tension}
  \varphi_0(x,y)=\hat \varphi_0(y)
\end{equation}
is independent of $x$, so that the $O(1)$ and $O(\epsilon^\alpha)$
terms of Eq.~(\ref{eq:EOM-parallel-ordered}) vanish. The leading order
in this equation could therefore be either $O(\epsilon^{2\alpha})$ or
$O(\epsilon)$. With Eq.~(\ref{eq:x-indep-tension}) we can solve the
$O(\epsilon^{1/2})$ part of Eq.~(\ref{eq:EOM-perp-ordered}) for $\vec
h_0(x,y)$ in terms of Fourier modes of the variable $x$ along the
lines of Sec.~\ref{sec:ampelfaktor} and use the result to evaluate
$X_0(x,y)$. It then turns out that the first term in $X_0$ implies
that $\varphi_1$ would have to grow without bound with increasing
system size (secular term), if the $O(\epsilon)$ terms alone were
required to cancel each other. However, the non--linear term can also
be balanced by the $O(\epsilon^{2\alpha})$ term after choosing
$\alpha=1/2$; i.e.\ the exponent $\alpha$ is fixed such that the
expansion coefficient $\varphi_1$ remains bounded \footnote{The small
  parameter $\epsilon^\alpha=\epsilon^{1/2}$ appearing here is the
  same as in the length scale separation
  Eq.~(\ref{eq:length-scale-sep}) observed in Sec.~\ref{sec:adi}.}.
The equation fixing $\varphi_1$ then reads
\begin{equation}
  \label{eq:phi1} 
 \partial_x^2\varphi_1(x,y)+\partial_y^2\hat
  \varphi_0(y)=X_0(x,y) \; .
\end{equation}
The balance of the secular terms implies the balance of the
$x-$averages of their derivatives that appear in Eq.~(\ref{eq:phi1}),
where $x-$averaging is defined by
\begin{equation}\label{eq:xavg}
\avg{g(x,y)}_x(y)=\lim_{l\to\infty} \mint{\frac{dx}{l}}{0}{l}g(x,y)\;.
\end{equation}
Note that $x-$averages of terms that are total derivatives of bounded
(non--secular) quantities with respect to $x$ all vanish upon formally
taking the coarse--graining length $l\to\infty$ in
Eq.~(\ref{eq:xavg}), so that we are left with
\begin{equation}
  \label{eq:cgapp} \partial_y^2\hat
  \varphi_0(y)=\frac{\hat\zeta}{2} \avg{ \partial_\tau (\partial_x
  h_0)^2}_x (y) \;.
\end{equation}
For the finite rod under consideration, the limit $l\to\infty$ is not
to be taken literally, though. Rather the average in
Eq.~(\ref{eq:xavg}) is required to become independent of $l$ to
leading order in $\epsilon$ for $l$ much smaller than the system size.
For the quantities of interest this was already established in
Sec.~\ref{sec:adiabatic_approximation}. Therfore, we can identify
Eq.~(\ref{eq:cgapp}) with the coarse--grained equation
Eq.~(\ref{eq:coarse-graining}). Relating corresponding quantities,
$\hat \varphi_0(y)$ that only depends on the slow variable $y$ is
recognized as the former coarse--grained tension $\varphi_l(s)$, while
the $x-$averaged expansion coefficient
$\avg{\varphi_{1}(x,y)}_x=\hat\varphi_{1}(y)$ corresponds to the time
derivative $\partial_\tau\epsilon_l(s)$ of the coarse--grained stored
length.


\end{document}